\documentclass[twocolumn]{aastex631}
\usepackage{amsmath}
\usepackage{longtable}
\newcommand{\phe}{\texttt{phot\_bp\_rp\_excess\_factor}}
\newcommand{\ago}{\texttt{astrometric\_gof\_al}}
\newcommand{\eb}{\begin{equation}}
\newcommand{\ee}{\end{equation}}

\newcommand{\uasyr}{$\mu$as yr$^{-1}$}
\newcommand{\gbp}{$G_{\rm BP}$}
\newcommand{\grp}{$G_{\rm RP}$}
\definecolor{rkka}{RGB}{219,66,32}

\shorttitle{Kinematic Distortions of the High-Redshift Universe}
\shortauthors{Makarov}
\journalinfo{Author's version of the paper with the same title published in NatAst, July 2025.\\
\vspace{20mm}\today}
\begin{document}

\title{Kinematic Distortions of the High-Redshift Universe as Seen from Quasar Proper Motions}

\correspondingauthor{Valeri V. Makarov}
\email{valeri.makarov@gmail.com}

\author[0000-0003-2336-7887]{Valeri V. Makarov}
\affiliation{U.S. Naval Observatory, 3450 Massachusetts Ave NW, Washington, DC 20392-5420, USA}

\begin{abstract}
Advances in optical astrometry allow us to infer the non-radial kinematic structure of the Universe directly from observations. Here I use a supervised machine learning neural network method to predict 1.57 million redshifts based on several photometric and metadata classifier parameters from the unWISE mid-infrared database and from Gaia. These estimates are used to divide the sample into three redshift bins: 1–2, 2–3 and  $>3$. For each subset, all available Gaia proper motions are used in a global vector spherical harmonic solution to degree 3 (30 fitting vector functions). I find significant differences in a few fitted proper motion patterns at different redshifts. The largest signals are seen in the comparison of the vector spherical harmonic fits for the 1–2 and 2–3 redshift bins. The significant harmonics include a rigid spin, a dipole glide from the north Galactic pole to the south and an additional quadrupole distortion. Validation tests with filtered subsamples indicate that the detected effect can be caused by hidden systematic errors in astrometry. The results are verified by using an independent source of redshifts and computing the observer’s Galactocentric acceleration. This study offers a new observational test of alternative cosmological models.
\end{abstract}

\section{Introduction} \label{intr.sec}
It is generally assumed that the Hubble flow of the expanding Universe has a time-dependent radial component but no time-dependent tangential component. The existence of such a non-radial component would imply that observers located at different parts of the Universe at earlier cosmological epochs would see an asymmetric distribution of radial expansion on the celestial sphere. This can be taken as an observable violation of the cosmological principle, extended to the time domain, which asserts that the observed motions of distant quasars do not exhibit a position dependence beyond that induced by the local effects of relativistic aberration and Doppler boosting. Using statistical and astrometric terms, the testable null hypothesis based on this principle is that a sufficiently large sample of quasars in a sufficiently large area of the sky is expected to have zero systemic tangential motion. Precision astrometric data for a large number of sources is used in this paper to directly explore the global pattern of proper motions of quasars separated by cosmological redshift.

The European Space Agency's Gaia mission \citep{2016A&A...595A...1G} delivered a trove of information about quasars and distant AGNs. The astrometric model used in the latest Gaia Data Release 3 \citep{2022arXiv220800211G} includes linear proper motions, which were estimated for millions of extragalactic sources at the sub-milliarcsecond level of precision. By design, the proper motion determinations for carefully selected quasars and galactic cores have been used to nullify the spin of the Gaia reference frame \citep{2018A&A...616A..14G}. Thus, the celestial reference frame (CRF) established by Gaia is predicated on the cosmological principle as a Bayesian prior, postulating that the system of distant quasars does not rotate as a whole. The measured proper motion values are believed to be mostly random observational errors, because the true angular motion of such distant sources is expected to be vanishingly small. The peculiar physical motion is also believed to be random and isotropically distributed. Any departures from a random proper motion field can be considered as a challenge to the accepted model, unless they can be explained as sky-correlated systematic errors of instrumental origin, or other known physical effects of non-cosmological nature.

The main purpose of this study is to test if large-scale systemic patterns of tangential motion can be detected at the early stages of cosmic expansion with the data available today. This study extends the approach presented by \citet{2022ApJ...927L...4M} taken from a relatively small patch of the sky to the global scale. 
The observed proper motion vector field of distant quasars $\boldsymbol{\mu}$ can be split into four
components:
\eb 
\boldsymbol{\mu}(l,b)=\boldsymbol{\Xi}(l,b)+\boldsymbol{X}(l,b)+\boldsymbol{\epsilon}+\boldsymbol{e},
\label{Xi.eq}
\ee 
where $l$ and $b$ are angular coordinates in the chosen celestial coordinate system, $\boldsymbol{e}$
is the random measurement error, $\boldsymbol{\epsilon}$ is the object-specific peculiar tangential motion,
$\boldsymbol{X}$ is the instrumental systematic error, and $\boldsymbol{\Xi}$ is the physical (cosmological) field representing possible departures from the isotropic radial pattern. The object-specific terms $\boldsymbol{e}$ and $\boldsymbol{\epsilon}$ are not functions of coordinates, and should be assumed to be
independent and uncorrelated on a large set of sources. The contribution from $\boldsymbol{\epsilon}$ can be safely ignored for this study, because with observed peculiar velocities of galaxies mostly within 1000 km s$^{-1}$,
the resulting peculiar proper motion is less than $0.16$ \uasyr\ at redshift $z=0.5$, which is much smaller than the
redshifts considered in this article. The random measurement errors $\boldsymbol{e}$ is the major limiting
factor, being two to three orders of magnitude greater per source. A large-number sample is needed to
statistically reduce this term to acceptable levels comparable to the possible signal $\boldsymbol{\Xi}(l,b)$.
Furthermore, it is not possible to separate the coordinate-dependent components $\boldsymbol{\Xi}$ and
$\boldsymbol{X}$. The tested hypothesis is that the cosmological signal $\boldsymbol{\Xi}$, unlike the
artificial systematic error $\boldsymbol{X}$, is also redshift-dependent, i.e., $\boldsymbol{\Xi}=\boldsymbol{\Xi}(l,b,z)$, which opens the possibility to detect it differentially.

In lieu of the accurate spectroscopically determined redshifts, which are only available in the footprint of the Sloan Digital Sky Survey (SDSS), synthetic redshifts are generated using the neural network method of supervised machine learning. The neural network is trained on a subset of 0.28 million quasars with SDSS redshifts extracted from the SDSS DR16Q version (catalog VII/289 in Vizier), and applied to 1.57 million Gaia CRF sources using mid-infrared fluxes, optical magnitudes, and Gaia ancillary parameters as prediction classifiers. The objects are divided into three non-overlapping bins by the synthetic redshift. Utilizing the large number of objects, a weighted least-squares fit with vector spherical harmonic (VSH), complete to degree 3, of the global proper motion field is calculated for each subset and for the entire sample of quasars. The statistically strong signals in the latter can be physical (component $\boldsymbol{\Xi}$ in Eq. \ref{Xi.eq}), but may also be considered as artifacts of the mission (component $\boldsymbol{X}$ in Eq. \ref{Xi.eq}). The significant differences between the three redshift subsets may indicate the presence of large-scale distortions of the kinematic structure of the Universe, unless they are related to implicit correlations with global calibration errors in Gaia DR3.

\section{Astrometric Quasar Sample}

This analysis begins with a sample of 1.6 million sources listed in Gaia DR3 \citep{2016A&A...595A...1G, 2022arXiv220800211G} and its annex Gaia Celestial Reference Frame (CRF-3) catalog \citep{2022A&A...667A.148G}. This initial sample includes $308,608$ cross-references to the Sloan Digital Sky Survey (SDSS). Matching the source Gaia DR3 main catalog to the SDSSDR16Q catalog \citep{2020ApJS..250....8L} in the Vizier database (table VII/289) retrieves only $297,379$ matches because of the earlier version of SDSS used for CRF-3. Furthermore, matching the original CRF-3 table with this cross-matched sample further reduces its size to $286,988$ common objects. This subset of Gaia CRF is used for training and verifying the Machine-Learning (ML) prediction of redshifts. Separately, the Gaia CRF-3 catalog was cross-matched with the unWISE catalog by \citep[][Vizier table II/363/unwise]{2019ApJS..240...30S} resulting in $1,569,472$ cross-matched objects. A master catalog was then generated with the relevant Gaia data supplemented with the unWISE identifications, MIR fluxes FW1 and FW2, their formal errors and quality indicators, SDSS designations and spectroscopic redshifts, when available.

Stars are present in the Gaia CRF-3 sample \citep{2024ApJS..274...27M} representing a potential risk for this study. Even the most distant stars in the outskirts of the Galaxy or in nearby galaxies are involved in coherent proper motion flows, which have no cosmological relevance. To estimate the rate of contamination by stars, the entire collection of CRF sources with synthetic redshifts used in this paper was cross-matched with the SDSS spectroscopic catalog. Within the main footprint of SDSS, 94\% of CRF sources brighter than $G=18.5$ mag have SDSS/BOSS spectra, and the fraction of entries spectroscopically classified as stars is 0.16\% (N. Secrest 2024, priv. comm.). The statistics are even better for the stripe 82 area, which is deeper (98.4\% completeness to $G < 19.5$ mag, 0.09\% contamination). Such a small admixture can hardly be a problem. However, the rate of contamination may be higher for fainter CRF sources, and in areas closer to the Galactic plane, as well as in the vicinity of Local Group galaxies. In total, 75\% of CRF sources are fainter than 19.5 mag where the rate of stellar contamination was estimated to be within 2\% by \citet{2022A&A...667A.148G}.

\section{Computations of proper motion fields}
\label{vsh.sec}

The most rigorous method of representing a systematic pattern in an observed proper motion field is to fit a sufficiently large set of vector spherical harmonic (VSH) functions (for a detailed description, see Sect. 6.1 and Suppl. Materials). The observed quasar proper motion field is approximated as
\eb 
\begin{split}\boldsymbol{\mu}(l,b)=&
\sum_{l=1}^{L}\sum_{k=0}^2 \sum_{m=1}^l [ c_{klm}{\bf EVSH}_{klm}(l,b)\\ & +
d_{klm}{\bf MVSH}_{klm}(l,b)],
\end{split}
\label{fit.eq}
\ee
where ${\bf EVSH}$ and ${\bf MVSH}$ stand for the VSH functions of the electric (multipole) and magnetic (curl) kind, cf. the Supplemental Material and the nomenclature cross-reference in \citep{2024arXiv241006075M}.
The limiting degree $L=3$, corresponding to 30 VSH fitting functions, and $(l,b)$ are the Galactic coordinates.
The fit captures only the smooth, systematic variation of $\boldsymbol{\mu}$ on the sky with a characteristic angular scale of $\sim 60\degr$. It is sensitive to the $\boldsymbol{\Xi}$ and $\boldsymbol{X}$ components
of the observed field, with a much reduced or negligible sensitivity to the random components. The VSH decomposition was computed several times for different samples of Gaia CRF sources. The general fit includes
all 1.2 million quasars with redshifts above 1.

\begin{figure}
    \includegraphics[width=0.45\textwidth]{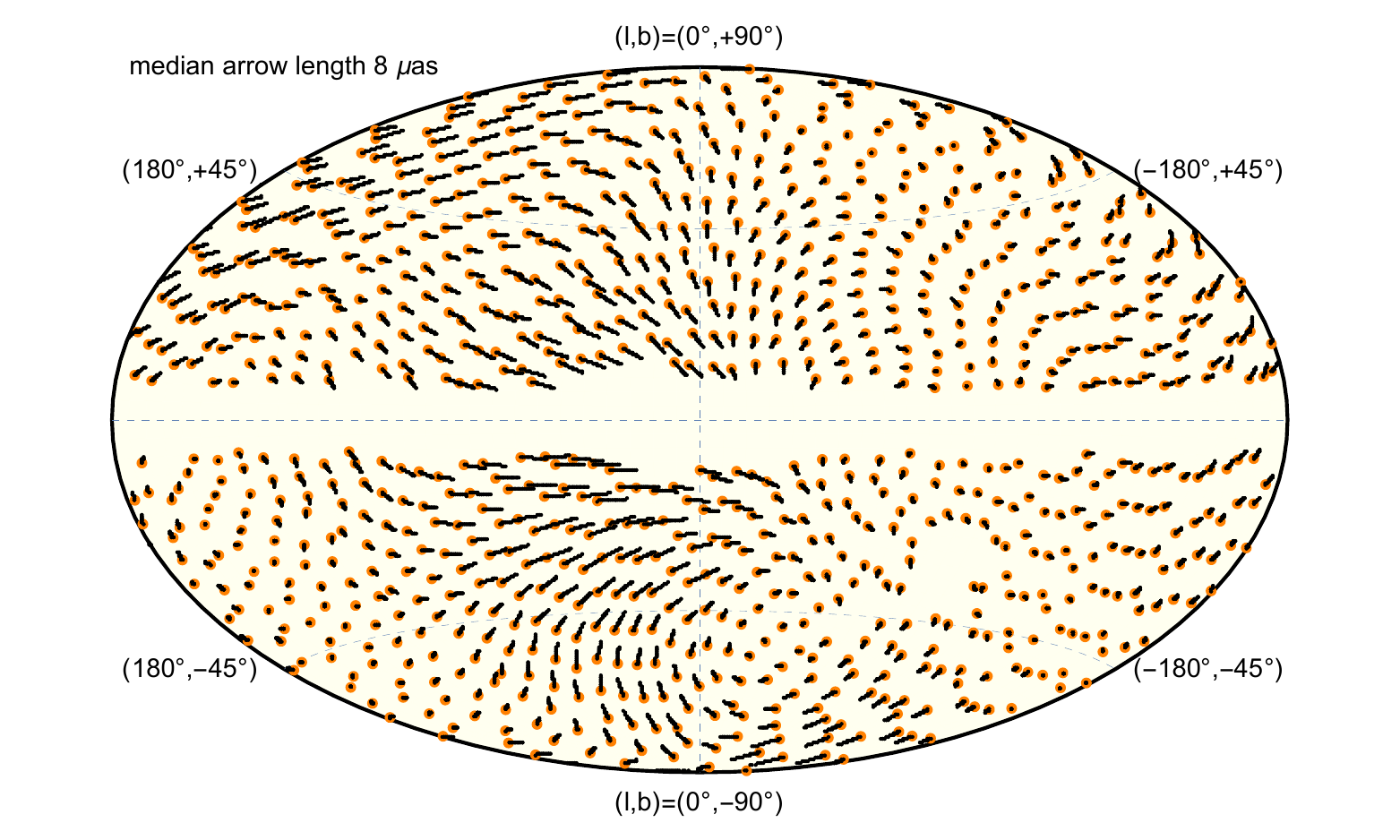}
   \caption{General VSH-fitted proper motion field of Gaia CRF quasars with ML-predicted redshifts $z>1$ on the celestial sphere.\\ Graphical presentation in the Aitoff Galactic projection with the Galactic center direction at the center of the plot. Orange dots at the origin of vectors indicate the mean positions of sources within the averaging cells. The length and direction of small vectors represent the cumulative VSH fit for the corresponding point. The median length of the arrows is 8 \uasyr.}
    \label{field0.fig}
\end{figure}

\begin{figure}
    \includegraphics[width=0.45\textwidth]{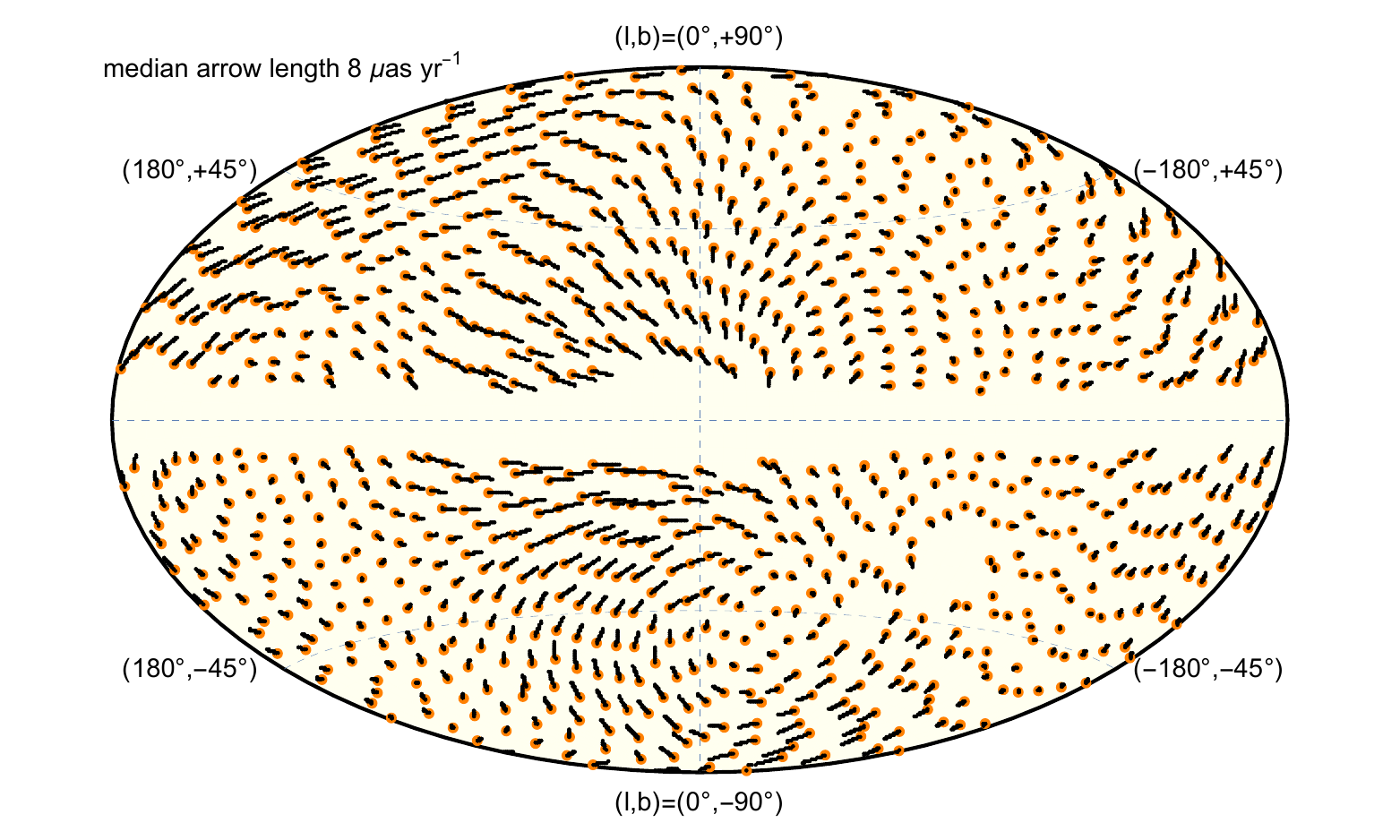}
   \caption{VSH-fitted Gaia CRF proper motion field of the subset of Gaia CRF quasars with redshifts $1<z<2$ on the celestial sphere. \\
The same graphical presentation is used as in Fig. 1.  The median length of the arrows is 8 \uasyr.}
    \label{field1.fig}
\end{figure}

\begin{figure}
    \includegraphics[width=0.45\textwidth]{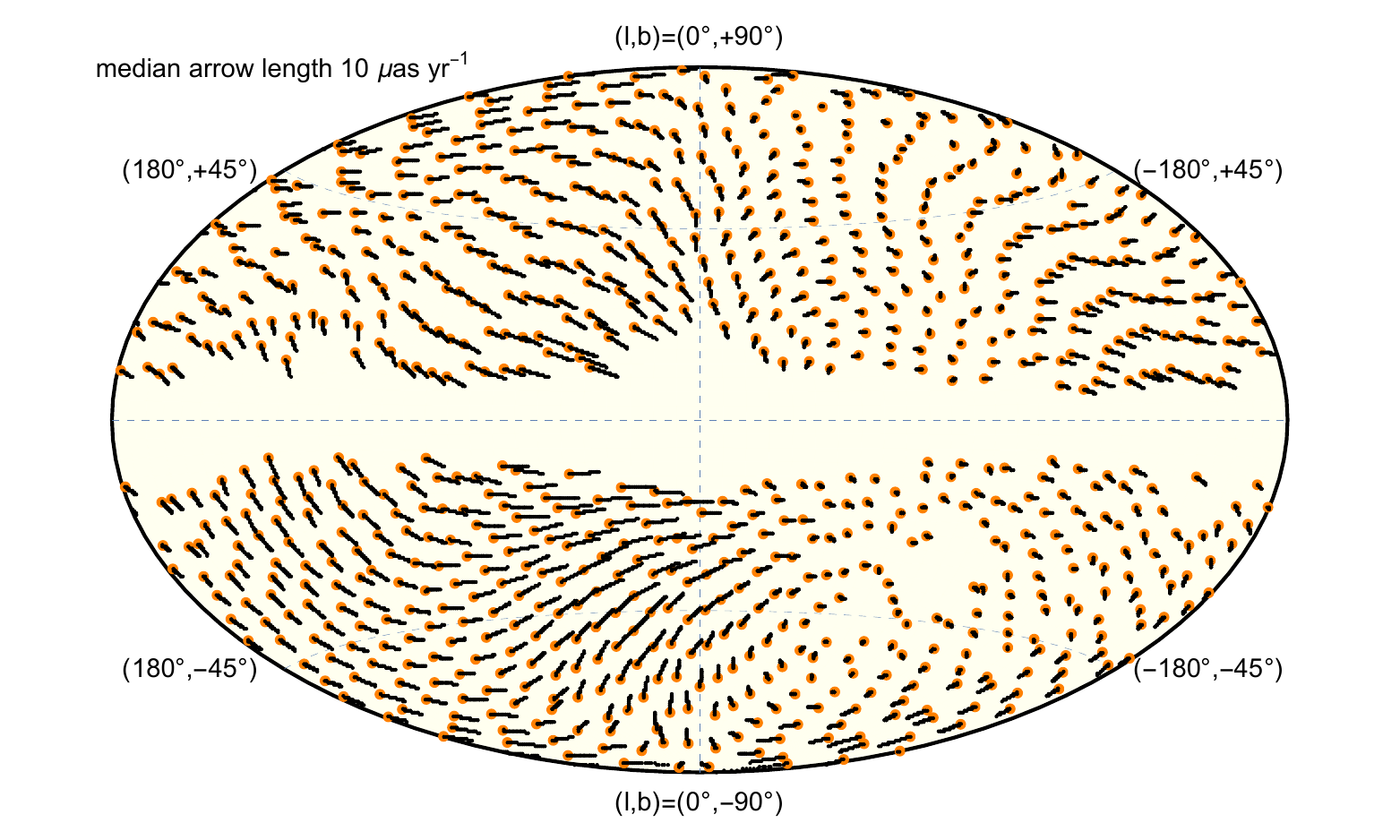}
   \caption{VSH-fitted Gaia CRF proper motion field the subset of Gaia CRF quasars with  redshifts $2<z<3$ on the celestial sphere.\\
   The same graphical presentation is used as in Fig. 1. The median length of the arrows is 10 \uasyr.}
    \label{field2.fig}
\end{figure}

The same least-squares fitting procedure is repeated for three subsets of the general sample grouped by redshift. The intervals of $z$, the median values, and the number of sources are specified in Table \ref{bins.tab}. Only the CRF quasars with ML-predicted redshifts above 1 are used in this study. The relatively nearby sources at lower redshifts are excluded because their astrometric quality is degraded in Gaia, and the VSH fits may be affected by the local kinematic perturbations \citep{2014A&A...561A..97P}, such as the nearest superclusters of galaxies and voids. This places the horizon of the current study well outside the Hubble distance. The least-squares solution for each sample produces 30 VSH coefficients $a_{j}$ corresponding to ordered coefficients $c_{klm}$ and $d_{klm}$ in Eq. \ref{fit.eq} (the ordering schema is described in Sect. 6.1.5 of Appendix) and their formal errors $\sigma_{a_j}$ in units of \uasyr. These values are combined in formal signal-to-noise ratio (S/N) estimates, which are computed as $|a_j|/\sigma_{a_j}$. The main results are presented in Table \ref{vsh.tab}, where the VSH functions are more conveniently presented in their tuple notation, for example, \{mag,1,2,2\}={\bf MVSH}$_{122}$.

\begin{deluxetable}{lcccr} \label{bins.tab}
\tablehead{\colhead{bin} & \colhead{Min[$z$]} & \colhead{Max[$z$]} & \colhead{Median[$z$]} & \colhead{number of sources}}
\startdata
1 & 1.0 & $\infty$ & 1.73 & 1163558\\
2 & 1.0 & 2.0 & 1.54 & 768621\\
3 & 2.0 & 3.0 & 2.28 & 370970 \\
4 & 3.0 & $\infty$ & 3.27 & 23967 \\
\enddata
\caption{Separation of quasars by redshift.
Redshift boundaries, median values, and number of sources for the four subsets of CRF quasars.}
\end{deluxetable}

\begin{deluxetable*}{cr|cc|cc|cc|cc} \label{vsh.tab}
\tablehead{
\multicolumn{2}{c|}{} & \multicolumn{2}{c|}{$z>1$} &\multicolumn{2}{c|}{$1<z<2$} &
\multicolumn{2}{c|}{$2<z<3$} &\multicolumn{2}{c}{$z>3$} \\ \hline
\multicolumn{1}{c}{number} & \multicolumn{1}{c|}{VSH}  & \multicolumn{1}{c}{$a_j$} & \multicolumn{1}{c|}{S/N}
& \multicolumn{1}{c}{$a_j$} & \multicolumn{1}{c|}{S/N} & \multicolumn{1}{c}{$a_j$} & \multicolumn{1}{c|}{S/N} & 
\colhead{$a_j$} & \colhead{S/N}\\
\colhead{}   &  \multicolumn{1}{c|}{}   &   \colhead{$\mu$as yr$^{-1}$} & \multicolumn{1}{c|}{} & \colhead{$\mu$as yr$^{-1}$} & \multicolumn{1}{c|}{} & \colhead{$\mu$as yr$^{-1}$} &\multicolumn{1}{c|}{} & \colhead{$\mu$as yr$^{-1}$} & 
 \colhead{}  
 } 
\startdata
 1 & \{\text{mag},0,1,0\} & 14.94 & 15.96 & 12.40 & 11.49 & 20.70 & 10.56 & -22.53 & 1.10 \\
 2 & \{\text{mag},1,1,1\} & 1.09 & 1.16 & -0.24 & 0.21 & 5.11 & 2.82 & 21.88 & 1.42 \\
 3 & \{\text{mag},2,1,1\} & -5.70 & 4.62 & -8.76 & 6.08 & 3.75 & 1.50 & -0.71 & 0.03 \\
 4 & \{\text{ele},0,1,0\} & 0.93 & 1.02 & -0.78 & 0.74 & 6.01 & 3.15 & -12.05 & 0.71 \\
 5 & \{\text{ele},1,1,1\} & -15.59 & 14.23 & -14.39 & 11.22 & -18.24 & 8.30 & -82.12 & 4.10 \\
 6 & \{\text{ele},2,1,1\} & -2.56 & 2.68 & -2.63 & 2.31 & -2.04 & 1.12 & -2.41 & 0.25 \\
 7 & \{\text{mag},0,2,0\} & -0.68 & 1.71 & -1.04 & 2.19 & 0.25 & 0.33 & 0.69 & 0.12 \\
 8 & \{\text{mag},1,2,1\} & -3.59 & 6.24 & -3.96 & 5.76 & -2.40 & 2.22 & -1.99 & 0.25 \\
 9 & \{\text{mag},2,2,1\} & 1.41 & 2.31 & 1.02 & 1.43 & 2.83 & 2.35 & -29.32 & 2.62 \\
 10 & \{\text{ele},0,2,0\} & -1.00 & 2.69 & -1.05 & 2.36 & -0.74 & 1.04 & -9.26 & 1.86 \\
 11 & \{\text{ele},1,2,1\} & -1.87 & 2.74 & -2.48 & 3.11 & -0.36 & 0.26 & 3.48 & 0.31 \\
 12 & \{\text{ele},2,2,1\} & -0.94 & 1.64 & -0.88 & 1.30 & -1.20 & 1.12 & -7.81 & 1.26 \\
 13 & \{\text{mag},1,2,2\} & 1.17 & 1.96 & 1.15 & 1.65 & 0.99 & 0.85 & -1.82 & 0.17 \\
 14 & \{\text{mag},2,2,2\} & -0.53 & 0.72 & -0.94 & 1.09 & 0.44 & 0.29 & -6.47 & 0.52 \\
 15 & \{\text{ele},1,2,2\} & 7.54 & 10.85 & 7.16 & 8.87 & 8.37 & 5.89 & 31.94 & 2.82 \\
 16 & \{\text{ele},2,2,2\} & -0.43 & 0.69 & 0.49 & 0.67 & -3.35 & 2.76 & -3.38 & 0.46 \\
 17 & \{\text{mag},0,3,0\} & -0.29 & 0.88 & -0.51 & 1.32 & 0.20 & 0.31 & 6.42 & 1.46 \\
 18 & \{\text{mag},1,3,1\} & -4.94 & 12.74 & -5.19 & 11.28 & -4.22 & 5.62 & -16.69 & 3.62 \\
 19 & \{\text{mag},2,3,1\} & 0.78 & 2.00 & 0.84 & 1.81 & 0.63 & 0.84 & -3.86 & 0.71 \\
 20 & \{\text{ele},0,3,0\} & -0.97 & 3.25 & -0.75 & 2.12 & -1.61 & 2.74 & -3.40 & 0.85 \\
 21 & \{\text{ele},1,3,1\} & -0.31 & 0.80 & -0.05 & 0.12 & -0.70 & 0.94 & 2.99 & 0.58 \\
 22 & \{\text{ele},2,3,1\} & 0.55 & 1.51 & 0.95 & 2.17 & -0.37 & 0.52 & 2.83 & 0.76 \\
 23 & \{\text{mag},1,3,2\} & 2.28 & 5.69 & 2.29 & 4.80 & 2.30 & 3.00 & 4.00 & 0.75 \\
 24 & \{\text{mag},2,3,2\} & 1.27 & 2.80 & 1.14 & 2.13 & 1.87 & 2.07 & 1.19 & 0.18 \\
 25 & \{\text{ele},1,3,2\} & -0.58 & 1.24 & -0.62 & 1.12 & -0.37 & 0.39 & -2.80 & 0.45 \\
 26 & \{\text{ele},2,3,2\} & 0.83 & 2.02 & 0.99 & 2.01 & 0.57 & 0.72 & 2.20 & 0.50 \\
 27 & \{\text{mag},1,3,3\} & -0.84 & 1.73 & -0.94 & 1.66 & -0.44 & 0.45 & 0.61 & 0.09 \\
 28 & \{\text{mag},2,3,3\} & -1.25 & 2.51 & -1.10 & 1.90 & -1.56 & 1.55 & -4.32 & 0.57 \\
 29 & \{\text{ele},1,3,3\} & -1.56 & 3.22 & -1.42 & 2.51 & -2.24 & 2.29 & -9.80 & 1.41 \\
 30 & \{\text{ele},2,3,3\} & 1.36 & 2.82 & 1.28 & 2.29 & 0.64 & 0.66 & -0.53 & 0.10 \\
\enddata
\caption{VSH fits of Gaia CRF proper motion fields. The results of VSH decomposition
of the general sample and the subsets segregated by redshift, including the names of fitting functions, estimated coefficients, and their S/N values.}
\end{deluxetable*}

Columns 3 and 4 in Table \ref{vsh.tab} contain the fitted values $c_{klm}$ and $d_{klm}$, and their formal S/N ratios for the most precise and reliable solution over the entire sample of CRF objects with $z_{\rm pre}>1$, which includes 1.16 million sources. Note that the coefficients correspond to the VSH functions including their standard normalization coefficients. We find that the most confidently detected pattern is that represented by the VSH number 1, \{mag,0,1,0\} $= (-(1/2) \sqrt{3/\pi} \cos b, 0)$, which is a rigid spin of the CRF around the Galactic polar axis in the left-hand sense. The emergence of this term and the other statistically significant term \{mag,2,1,1\} is due to the fact that the alignment of rigid spin for Gaia DR3
was done separately from the overall VSH analysis based on the smaller preliminary CRF sample of quasars \citep{2022A&A...667A.148G}. The spin terms are free parameters for the Gaia mission in general, which cannot be determined intrinsically from its own proper motion measurements without a Bayesian prior. The corresponding rate of rotation is $-7.3 \pm 0.5$ \uasyr\ in the north pole direction. The sample of Gaia-selected candidate quasars includes a finite fraction of Galactic stars, which are involved in the galactic rotation, and the cut on redshift applied in this study may help to remove them. The second largest term is number 5,
\eb
\text{\{ele,1,1,1\}}=\left(\frac{1}{2} \sqrt{\frac{3}{2 \pi}} \sin l, \frac{1}{2} \sqrt{\frac{3}{2 \pi}} \cos l \sin{b}\right),
\ee
which is a gliding motion with an amplitude of $5.39\pm 0.38$ \uasyr\ toward the Galactic center. Unlike the previous magnetic terms, this is the only signal in the proper motion field of distant quasars that is expected. 
It represents the secular aberration drift pattern caused by the main component of the acceleration of the Solar system barycenter toward the Galactic center of mass \citep{2003A&A...404..743K, 2006AJ....131.1471K}. 
This estimate, corresponding to an absolute acceleration of $2.48\times 10^{-10}$ m s$^{-2}$, is within one formal error of the 
value obtained by \citet{2021A&A...649A...9G} on the entire sample of Gaia CRF sources: 
$(2.32 \pm 0.16)\;10^{-10}$ m s$^{-2}$.

The other significant VSH terms in columns 2 and 4, including the second-degree harmonics  \{mag,1,2,1\},  \{ele,1,2,2\} and the third-degree harmonics  \{mag,1,3,1\},  \{mag,1,3,2\}, do not find a specific explanation. As discussed in the next section, they may be caused by a very strong primordial gravitational wave propagating through the local epoch of the Universe, or by intrinsic distortions of the isotropic Hubble flow, in principle. The Occam's razor principle dictates, however, that the more mundane explanation should be accepted, that these unexpected patterns are caused by hidden systematic errors in Gaia observations. Fig. \ref{field0.fig} shows the combined proper motion field represented by the VSH fit for the general sample. The presence of the secular aberration signal is marginally visible only in the first and second Galactic quadrants ($0\degr < l < 180\degr$). The effect is completely washed out in the other hemisphere, where a vortex at $b\simeq -45\degr$ close to the principal meridian is the most conspicuous feature. The interference with the higher-degree signals generates a striking asymmetry between these hemispheres, as most of the proper motion field appears to be in the first and second quadrants. The enigmatic \{mag,1,3,1\} terms is the third largest in terms of statistical confidence (S/N$=12.7$). It generates two symmetric vortexes in the prime meridian at $b=-45\degr$ and $b=+45\degr$, but we can only see the southern one in Fig. \ref{field0.fig}, as the northern counterpart is washed out by the other harmonics. 

The same VSH decomposition is produced for a subset of 0.769 million CRF quasars with predicted redshifts in the interval $[1,2]$. The results are tabulated in Table \ref{vsh.tab}, columns 5 and 6, and the fitted proper motion field is shown in Fig. \ref{field1.fig}. The visual pattern is quite similar to that for the general sample in \ref{field0.fig}. We obtain practically the same set of significant signals with the leading \{mag,0,1,0\} term. The similarity is not surprising, because this subset of sources comprises the majority of the general sample, and have the highest weight there. The VSH coefficients are numerically close too. If we ignore the intrinsic positive correlation of these estimates and compute the formal uncertainty of the difference for each term as the quadratic sum of the formal errors, these differences are all statistically insignificant, with the largest normalized  offset at S/N$=1.78$ found for the same rotation term \{mag,0,1,0\}. 

Another VSH analysis is made for the proper motion field of quasars in the interval $z_{\rm pre}\in [2,3]$. The main results are presented in Table \ref{vsh.tab} (columns 7 and 8) and in Fig. \ref{field2.fig}. Because of the smaller number of sources in this redshift bin (0.371 million), the formal uncertainties of the VSH coefficients are considerably larger. The median magnitude of the fitted proper motion field increased from 8 \uasyr\ to 10 \uasyr, possibly because of the greater dispersion of the estimated coefficients. There are obvious differences with both the general sample and $z_{\rm pre}\in [1,2]$ results. The number of confidently detected signals is reduced to four, with \{mag,0,1,0\} still being the largest term. Its value, however, almost doubled for the more distant quasars. A much smaller increase in absolute value is also seen for the secular aberration term \{ele,1,1,1\}, implying a faster acceleration of the observer toward the Galactic center. The higher-degree terms \{ele,1,2,2\} and \{mag,1,3,1\} have barely changed, and the southern vortex below the Galactic center is still clearly present. The rigid rotation term \{mag,2,1,1\}, which is strong and persistent in the two previous fits, is now practically gone. The $z_{\rm pre}\in [1,2]$ and $z_{\rm pre}\in [2,3]$  determinations are statistically independent, because they are based on separate samples of sources. It is the difference between these two fits that has the potential of revealing cosmological distortions violating the cosmological principle. The difference of the VSH coefficients for each term divided by the combined formal error yields the S/N ratio of this signal. This calculation reveals that the only significant differences (S/N$>3$) are found for VSH terms of first degree \{mag,2,1,1\} (4.3), \{mag,0,1,0\} (3.7), and \{ele,0,1,0\} (3.1). The two magnetic terms represent rigid rotations of the entire proper motion field around the axis orthogonal to the direction of the Galactic center and around the Galactic poles, respectively. The corresponding relative spin rates between the two $z$-samples are 4.3 \uasyr\ and 4.1 \uasyr. Contrary to the \{mag,2,1,1\} term, the emergent \{ele,0,1,0\} term is not at all significant in the previous determinations for the closer quasars and the general sample. This term is, specifically,
\eb
\text{\{ele,0,1,0\}}=\left(0, \;-\frac{1}{2} \sqrt{\frac{3}{\pi}} \cos{b}\right),
\ee
representing a glide along the Galactic axis of rotation. The difference between the $z_{\rm pre}\in [2,3]$ and $z_{\rm pre}\in [1,2]$ terms corresponds to a glide with an amplitude of 3.3 \uasyr\ toward the south Galactic pole. We note that the two somewhat less significant signals are in the \{ele,2,2,2\} VSH term at S/N$=2.7$:
\eb
\begin{split}
&\text{\{ele,2,2,2\}}=\\
& \left(\frac{1}{2} \sqrt{\frac{15}{2 \pi}} \cos(2 l) \cos b, - \sqrt{\frac{15}{2 \pi}} \cos l \sin l \cos b \sin{b}\right),
\end{split}
\ee
and the third rotation term \{mag,1,1,1\} around the Galactic center at S/N$=2.5$.

A fourth VSH fit was performed for the sample of CRF quasars with ML-predicted redshifts above 3, but it obtained relatively uncertain results because of the much smaller number of available data vectors (0.024 million). The results are shown in Table \ref{vsh.tab}, columns 9 and 10. Two of the previously detected terms still appear with formal S/N above 3, viz., the secular aberration dipole \{ele,1,1,1\} and the enigmatic vortex \{mag,1,3,1\}, but the precision of these coefficients is much degraded. In comparison with the general sample, this fit does not reveal any statistically significant differences except for the dipole \{ele,1,1,1\}, where a formal S/N=3.3 is estimated. The apparent dependence of this signal, which is the only effect we expected to see, on redshift is one of the main results of this study.

Several VSH terms emerge as statistically significant signals with high S/N ratios, but the overall dispersion of observed proper motions is still dominated by the random or high-frequency component. This is seen from the computed reduced $\chi^2$ values of the cell-averaged proper motions (normalized with their formal covariances). The pre-fit values are 2.11, 1.95, 1.71, and 1.56 for the overall $z>1$, $z\in [1,2]$, $z\in [2,3]$, and $z>3$ samples, respectively. The post-fit residual values are, correspondingly, 1.96, 1.85, 1.64, and 1.52. Despite the moderate reduction of $\chi^2$ statistics, the formal $F$-test shows a zero $p$-value because of the large number of degrees of freedom.

 \section{Discussion and Conclusions}
 
The ML-predicted redshifts computed by a neural network method from several photometric and astrometric data types from unWISE and Gaia have limited precision for individual sources but are adequate for the main purpose, which was to roughly segregate the CRF sources by redshift. The leakage of Gaia CRF sources between the adjacent coarse bins of 1--2, 2--3, and above 3 is limited to several percent (see Section \ref{neu.sec} for detailed estimation on the SDSS training set). This residual mixing of sources can only diminish the differential signals in the VSH fits for each redshift bin and in comparison with the overall sample. 
Indeed, the null hypothesis to be tested is that the cosmic signal $\boldsymbol{\Xi}(l,b)$ in Eq. \ref{Xi.eq} is zero, in accordance with the cosmological principle and kinematically isotropic universe. The significant VSH terms for the general $z>1$ sample, which serves as the control sample, do not disprove the null hypothesis, because they can be interpreted as the instrumental systematic error $\boldsymbol{X}(l,b)$. Separating the general sample into three non-overlapping subsets allows us to find the differential signal by comparing the resulting VSH coefficients between the subsets and the control sample. 
%If the cosmic signal in a given VSH term is a smooth and integrable function of $z$, the binning operation is equivalent to applying a linear smoothing filter with a rectangular response function. Due to the limited bandwidth of that filter, the resulting convolution is ``redder", i.e., more flat than the true dependence on redshift. Randomly selecting sources from the general sample irrespective of their redshift produces statistically identical result to the control sample. A moderate degree of leakage between the adjacent $z$-bins is a case between these margins, because it introduces one additional convolution with a bell-shaped filter function similar to the one depicted in Fig. \ref{zpre.fig}. The result is flatter than the averaging with a rectangular linear filter, which is conservative. 
For verification of the main result, the entire computation was repeated using the spectroscopic and synthetic redshifts from the Quaia catalog by \citet{2024ApJ...964...69S}, which is an independent source of data (see Sect. 6.2.1).

This differential approach allows us to probe for possible redshift-dependent cosmological signals in the observed proper motion field of distant quasars. The proper motion fields were fitted with 30 VSH functions up to degree 3 in separate least-squares adjustments for each redshift bin, as well as for the entire sample of $1,163,558$ sources with synthetic redshifts above 1. The latter decomposition provides an estimate of possible artifacts of instrumental origin (i.e., false signals), while the deviations of individual fits for the redshift bins from the general field may include legitimate cosmological signals.

%The overall VSH fit reveals six terms with statistically significant coefficients (S/N$>4$). 
%The two largest terms in the general sample by absolute values and formal S/N ratios are $\{\text{mag},0,1,0\}=(-\frac{1}{2} \sqrt{\frac{3}{\pi}} \cos b, \;0)$ and $\{\text{ele},1,1,1\}=(\frac{1}{2} \sqrt{\frac{3}{2\pi}} \sin l, \;\frac{1}{2} \sqrt{\frac{3}{2\pi}} \cos l\,\sin b)$. The first one represents the rigid spin of the entire proper motion system around the north Galactic pole in the counter right-hand direction. The emergence of this rotation term, as well as the relatively less significant $\{\text{mag},2,1,1\}$ representing the rigid spin around the Galactic Y axis, is caused by the overall posterior adjustment of the overall DR3 proper motion system, which was done separately and on a different selection of candidate quasars. 
%A small admixture of stars contaminating the Gaia spectrophotometric sample causes a bias in the spin terms due to the differential Galactic rotation and warp. The Gaia CRF3 sample is cleaner with additional astrometric filters applied. We also note that the spin determination in the Gaia pipeline has been limited to only the three first-degree magnetic terms, and truncating the VSH decomposition so severely may affect the result because of the cross-correlations of VSH terms. 

The largest signal in the $\{\text{ele},1,1,1\}$ harmonic is truly the only place where we expected to find it. This term represents the main component of the secular Galactic aberration effect, which has been estimated by \citet{2021A&A...649A...9G} from the entire CRF sample. 
%As explained ibid., this determination is robust with respect to possible systematic errors from the magnitude and color distributions of CRF sources, a residual contamination from Galactic field stars, and other conceivable astrophysical and technical effects. 
The estimate for the Galactocentric component of the dipole pattern obtained for a truncated sample at redshifts above 1, $5.39\pm 0.38$ \uasyr, is within one formal standard deviation of their result. The presence of this dipole pattern in the graphical representation of the fitted proper motion field in Fig. \ref{field0.fig} is clearly seen only in the first and second quadrants, however. The visual impression of the strong asymmetry between the two hemispheres separated by the principal Galactic meridian is the result of an interference of the aberration dipole with other strong second- and third-degree VSH terms, which do not have an immediate explanation. The unexpected signals include a third-degree magnetic term, which is responsible for the conspicuous vortex south of the Galactic center. It is roughly coincident with the location of the Small Magellanic Cloud, but there are no obvious reasons of astrophysical or technical kind for this association.

The motion of the Solar system with respect to the local rest frame as determined from the CMB temperature distribution dipole (370 km s$^{-1}$) generates a reflex proper motion dipole \citep{1986AZh....63..845K} with an amplitude of 0.046 \uasyr\ of sources at $z=1$ and $z=2.53$, and an even smaller apparent motion for sources between these redshifts. Galaxy groups can have somewhat higher peculiar velocities relative to the Hubble flow on a smaller angular scale. These effects are roughly 20 times smaller than what can be measured today. An observable signal in quasar proper motion would likely be related to unexpectedly large perturbations of matter density in the Universe that propagated to later epochs or even increased in relative amplitude
\citep{1968Natur.217..511R}, and the related large-scale variations of gravitational potential via the Poisson equation. The manifestation of these early Universe perturbations in the CMB temperature map is known as the Sachs-Wolfe effect \citep{1967ApJ...147...73S} seen as a constant zero-point offset of the spatial spectral power. Unfortunately, the CMB determination is rather ambiguous because of the intrinsic uncertainty of the spectral power in the lowest degrees of spherical harmonics, where this effect would be most visibly present. Our investigation offers an alternative way of quantifying possible global patterns of quasar proper motions at different cosmological epochs. 

Within the Friedmann-Robertson-Walker metric, galaxy clusters and quasars are receding from the observer in a purely radial pattern that is invariant to the observer's location. The cosmic expansion pattern is strictly radial and independent of the observer's location. In the alternative radially inhomogeneous Lema{\^i}tre-Tolman-Bondi metrics, the cosmic expansion rate becomes not only time-dependent but also direction- and distance-dependent because of the offset of the observer from the center of the spherically symmetric expansion pattern. This generates a systemic pattern of tangential velocities, which is also called the ``cosmic parallax" in the literature \citep{2009PhRvL.102o1302Q}. The emerging global field of tangential motion is a function of the radial dependence of expansion rate and the position vector of the observer. This field can be represented by a set of VSH functions with redshift-dependent coefficients. Ideally, the complete 3D VSH decomposition could be used, but the limited volume and accuracy of the present-day data forces us to apply only crude separation of the quasar sample by mostly synthetic redshifts.

Possible large-scale inhomogeneities or ``distortions" of the quasar proper motion fields not violating the cosmological principle also include the primordial gravitational wave propagating through the local part of the universe \citep{1966ApJ...143..379K, 1996ApJ...465..566P,1997ApJ...485...87G}. The effect is mostly confined to the VSH terms of second degree (quadrupole terms) and it scales with the magnitude of the propagating wave, which is locally strongly dilated. At earlier cosmological epochs, however, the strength of primordial waves can be greater, and the effect may in principle be observable in higher degrees of the proper motion field. Although we do find a few formally significant second- and third-degree terms in the general fit, they appear to be remarkably consistent across the partitions by redshift. 

The strong systematic patterns in the general field, including the vortex motion south of the Galactic pole, can be attributed to yet unknown technical perturbations of the Gaia measurements. The astrometric calibration of the Gaia data is quite complex \citep{2021A&A...649A...2L}, involving a large number of nuisance parameters. The systematic part of the error budget is certainly correlated with source's brightness, color, the number of fitted parameters, deviation from the expected point-like image, optical variability, etc. Some of these factors can be indirectly correlated with the redshift. To shed more light on possible hidden systematics, a number of verification test has been performed with different filters on the CRF sample. It was found, in particular, that limiting the source sample to objects brighter than $G=19.8$ mag and removing the measured proper motions above $3\sigma$ results in a reduction of all prominent VSH coefficients by approximately 40\%, including the first-degree magnetic terms and the secular aberration dipole. Unfortunately, this filter also increases the formal errors of the VSH coefficients due to the much smaller number of accepted data points, and the result remains somewhat inconclusive. Additional tests were performed to estimate the sensitivity of the VSH fit to a small fraction of cells with statistically outlying average proper motions. Filtering out 37 partitions (2\% of the original set) with proper motion $\chi^2$ values above 15.7 was found to change the estimated VSH coefficients mostly within $1\sigma$ (formal error) confirming that the detected pattern is a global field rather than a feature limited to a fraction of deviant locations.

The significant differences in VSH terms between the 1--2 and 2--3 redshift bins (with the same selection filters) are harder to explain. These differences are seen only in first-order VSH terms, while the higher-order terms are quite consistent between the two samples, including the southern vortex. This persistent feature appears to be aligned with the location of the Small Magellanic Cloud, but it is not obvious how this dwarf galaxy could generate a persistent instrumental effect of these character and scale. Two rigid rotation terms and the Galactocentric dipole term differences have formal S/N ratios above 3. The rotation terms differ by approximately $4\pm1$ \uasyr. Formally, this result implies that the universe has a net rotation as a whole, which is different at different cosmological epochs. The sign of the implicit angular acceleration cannot be obtained by the astrometric method  because of the interfering technical alignment of the Gaia CRF with ICRF3. The apparent dependence of the Galactocentric dipole component on redshift may seem even more baffling, because the aberration of light effect is distance-invariant. Dipole cosmology models have been proposed \citep{2023arXiv230516177E}, which incorporate uniform flows of radiation and matter that can be time-dependent. The relative flow of matter is detectable within these models as a redshift-dependent dipole of the proper motion field. Bianchi Type I universes can, in principle, be consistent with global velocity dipoles through an anisotropic metric of the form $ds^2 = -dt^2 + a_x^2(t) dx^2 + a_y^2(t) dy^2 + a_z^2(t) dz^2$ \citep[][and references therein]{2014Galax...2...22P, 2022NewAR..9501659P}. These models predict time-dependent bulk flows on a grander scale than has been tested before.

\section{Methods}

\subsection{Neural network training and prediction of redshifts for 1.5 million Gaia CRF sources}
\label{neu.sec}

\citet{2023ApJS..264....4M} synthesized 0.617 million redshifts for an intersection of the Gaia DR3 catalog and MIRAGN, the infrared photometry-based catalog of optical quasars \citep{2015ApJS..221...12S}. These ML-predicted redshifts were used as secondary data in the construction of the Candidate Double AGNs catalog to filter out the closer objects at $z>0.5$. Based on this experience, one of the best performing ML techniques is chosen here to generate a larger number of synthetic redshifts, viz., the neural network method. This method belongs to the class of supervised machine learning techniques. The neural net is trained to predict redshifts from a set of input vectors called classifiers on a subset of data with known correct answers. In this case, the training set comprises $277,472$ examples, i.e., objects with all the required classifiers available in the master catalog, and the accurate SDSS spectroscopic redshifts \citep{2020ApJS..250....8L}. The set of classifiers was selected after an extensive search across the entire parameter space in Gaia and unWISE \citep{2019ApJS..240...30S}. Ideally, a classifier is a parameter, for which a strong and preferably monotonical dependence on redshift is observed. \citet{2023ApJS..264....4M} used four photometric classifiers: MIR colors $W1–W2$ and $W2–W3$ from WISE, the mixed color $G–W1$, and the Gaia determined color $G_{\rm BP}–G_{\rm RP}$. The ancillary parameter \phe\ from Gaia \citep{2021A&A...649A...5F} was found to be well correlated with redshift, and it is employed in this analysis too. This parameter of photometric nature is not related to the astrometric proper motion, but quantifies how well the observed images of each source match the expected point-source response function in the two filters at the pixel level.

\begin{figure}
    \includegraphics[width=0.40\textwidth]{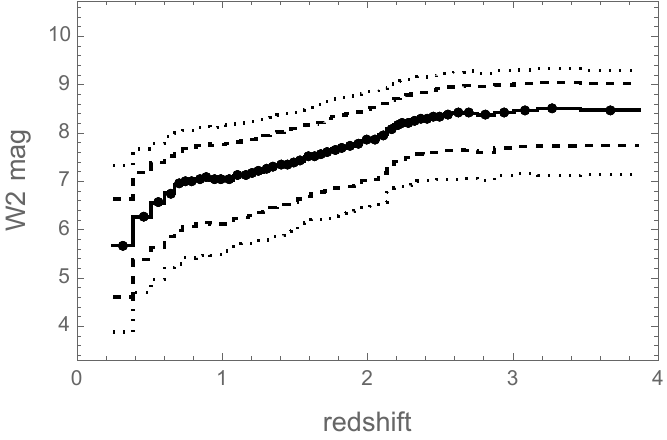}
   \caption{Median infrared magnitude versus redshift.\\
   Statistical dependence of instrumental $W2$ magnitude computed from unWISE flux $FW2$ with an arbitrary zero-point on spectroscopic redshift.
    The broken solid line with dots shows the median values in 50 equal bins of sorted redshifts. Each dot  in this line indicates the median value of redshift for the corresponding bin, which is not centered within the bin. The dashed lines below and above the median curve show the 0.16 and 0.84 quantiles, respectively, of the sample distribution, which provide robust standard deviation estimates. The outer dotted lines are the boundaries of the wider 0.05--0.95 quantile range.}
    \label{w2.fig}
\end{figure}

The other change concerns the source of MIR photometry. As Gaia CRF contains a much larger number of optically faint sources compared with MIRAGN, the unWISE catalog is used instead of WISE, because it goes significantly deeper. This change comes at the cost of loosing the $W2-W3$ color, which is an efficient redshift classifier. To compensate for this loss, an instrumental magnitude $W2$ computed from the tabulated flux $FW2$ is included as a separate classifier. Fig. \ref{w2.fig} shows the statistical dependence of this classifier on redshift and its robustly estimated uncertainty boundaries. Unfortunately, this dependence is adequately discriminating only at smaller redshifts, and becomes quite flat at $z>2$. The same can be said about the Gaia \ago\ parameter, which works well only at $z<0.5$ \citep[cf. Fig. 4 in ][]{2023ApJS..264....4M}. The optical color \gbp$-$\grp\ is somewhat sensitive to redshift at the high end of $z$, but it is not monotonical and much dispersed there due to the faintness of the bulk of CRF objects \citep[Fig. 2 in ][]{2023ApJS..264....4M}. Six independent or partially independent classifiers were chosen for the neural network training: the instrumental $W2$ magnitude, $W1-W2$, $G-W1$, and \gbp$-$\grp\ colors, and the Gaia ancillary parameters \phe\ and \ago.

\begin{figure*}
    \includegraphics[width=0.40\textwidth]{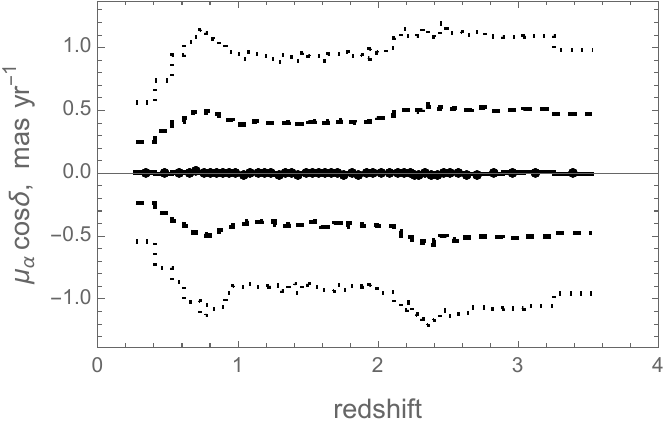}
    \includegraphics[width=0.40\textwidth]{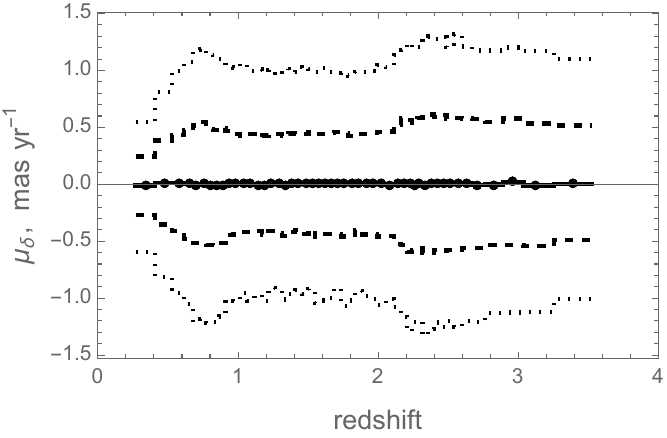}
   \caption{Median Gaia proper motions versus redshift.\\
   Statistical dependence of proper motion components in RA (left plot) and Decl (right plot) on spectroscopic redshift.
    The broken solid line with dots shows the median values in 50 equal bins of sorted redshifts. Each dot  in this line indicates the median value of redshift for the corresponding bin, which is not centered within the bin. The dashed lines below and above the median curve show the 0.16 and 0.84 quantiles, respectively, of the sample distribution, which provide robust standard deviation estimates. The outer dotted lines are the boundaries of the wider 0.05--0.95 quantile range.}
    \label{muz.fig}
\end{figure*}

Gaia-measured proper motions is the main type of data used in this study, and its systematic statistical independence of redshift is to be verified. This is done by producing similar plots to Fig. \ref{w2.fig} for proper motions binned by redshift, as well as for the grand CRF sample. The median proper motion as a function of redshift shown in Fig. \ref{muz.fig} is close to zero in both RA and Decl. components, as expected. The robust standard deviation intervals defined by the 0.16 and 0.84 quantiles of the sample distributions are not obviously widening with $z$ in the range of interest for this study. The observed dispersion of proper motions shows a distinct bump at $z$ between 0.8 and 1.0, although the quasars are not statistically fainter at this redshift. The origin of this feature is not known to me, but the important conclusion is the absence of global correlation of Gaia-measured proper motions with redshift.

\begin{figure}
    \includegraphics[width=0.5\textwidth]{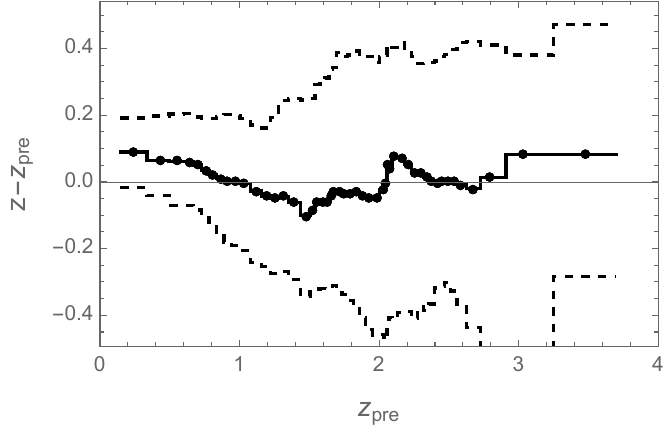}
   \caption{Performance of Machine-Learning prediction of redshifts.\\
   Statistical dependence of differences between ML-predicted redshifts $z_{\rm pre}$ and observed spectroscopic redshifts $z_{\rm obs}$ for the training set of 0.28 million sources in the training sample.
    The broken solid line with dots shows the median values in 50 equal bins of sorted predicted redshifts. Each dot indicates the median value of redshift for the corresponding bin, which is not centered within the bin. The dashed lines below and above the median curve show the 0.16 and 0.84 quantiles, respectively, of the sample distribution, which provide robust standard deviation estimates.} 
    \label{z-z.fig}
\end{figure}

Construction and training of the neural network on this set of classifiers revealed a partial degradation of performance  compared with a similar ML-prediction performed by \citet{2022ApJ...927L...4M} on a sample of MIRAGN quasars, attributed to the missing $W2-W3$ classifier and the inclusion of fainter, less precisely measured sources. This is confirmed by Fig. \ref{z-z.fig}, where the median error of predicted redshifts and its robust standard deviation interval is plotted as a function of the predicted redshift $z_{\rm pre}$. The dispersion of differences greatly increases from 0.1 at the low end to approximately 0.4 at the high end of predicted redshifts. Besides this underperformance at higher $z$, we observe a strong and variable bias of $z_{\rm pre}$, which is probably related to asymmetries in the distributions of classifiers with respect to the means at different redshifts. This bias is, however correctible, and it has been taken out of the general sample of predicted redshifts.

\begin{figure}
    \includegraphics[width=\columnwidth]{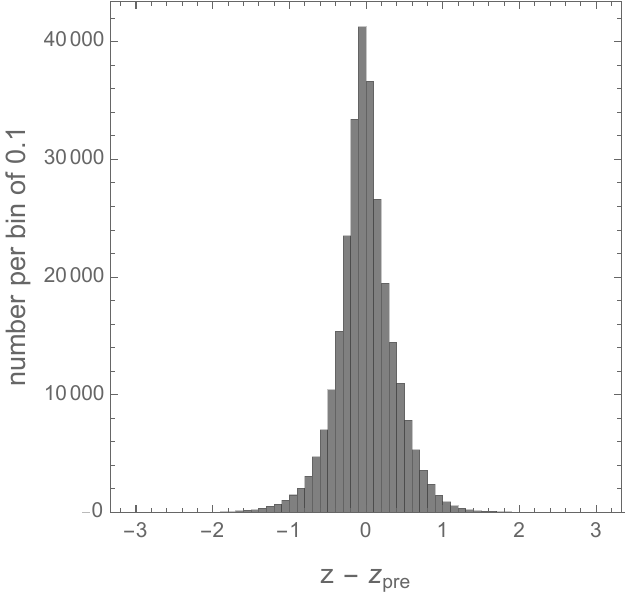}
    \caption{Errors of ML-predicted redshift.\\
    Histogram of ML-generated redshift deviations $z-z_{\rm pre}$ for 0.28 million quasars in SDSS obtained with the neural network regression method}.
    \label{zpre.fig}
\end{figure}

The trained neural network was then utilized on the general sample of Gaia+unWISE sources and predicted redshifts were generated for $1,568,693$ sources. The previously detected bias was added to the synthesized values using a linear interpolation function of zeroth order. The distribution of the differences between the ML-generated and SDSS-observed redshifts is shown in Fig. \ref{zpre.fig}. 
To verify the dispersion levels of the synthetic redshifts, an additional cross-validation was performed using the standard 4-fold cross-validation technique. The sample of SDSS sources (the entire training set) was randomly divided into four equal parts. For each of these subsets, a separate neural network was generated (trained) using the union of the remaining three subsets. Each trained network was applied to the reserved quarter, and the predicted redshifts were combined from the four networks and compared with the observed redshifts. The results were practically indistinguishable from the overall residual statistics shown in Figs. \ref{zpre.fig} and \ref{z-z.fig}. The test revealed that 31\% of the training sample has prediction errors within 0.1, 54\% within 0.2, and 98\% within 1.

In the context of this study, the performance of ML-regression can be quantified in two ways. As shown in Fig. \ref{z-z.fig}, the robust standard deviation intervals can be computed for each bin of $z$ on the training set. We note, however, that these individual quantities are not used in the present study. Instead, the entire collection of CRF sources is divided into three wide bins of $z$ using the predicted values. Therefore, the amount of leakage between the $z$-bins is of greater interest (Section \ref{neu.sec}). Swapping sources between adjacent bins is conservative in the context of this study, because it can only reduce the redshift-dependent differential signals of the corresponding proper motion fields. The ML method is found inefficient in detecting very high redshifts, which is caused by the paucity of such quasars and their indiscriminate properties. It is of little consequence for the present analysis, which is effectively limited to $z_{\rm pre}<3$. 

\subsection{Vector spherical decomposition of tangential vector fields}
\label{deco.sec}
While scalar spherical harmonics have been used in other astronomical applications, mostly to quantify the systematic differences of parallaxes and coordinate components between astrometric catalogs, the VSH decomposition of astrometric data is a relatively novel technique, which was pioneered by \citet{1990srst.conf..151M} and \citet{2004ASPC..316..230V}, and used to study the local Galaxy kinematics \citep{2007AJ....134..367M, 2015AJ....149..129M}, reveal systematic differences between global astrometric catalogs \citep{2011AstL...37..874V, 2012A&A...547A..59M}, determine the acceleration of the Solar system in the Galaxy \citep{2011A&A...529A..91T}, and to determine the link between the radio and optical fundamental reference frames \citep{2016A&A...595A...5M}. 

The implementation of the VSH analysis in this study closely follows that in \citep{2022AJ....164..157M}. A very detailed description can also be found in Suppl. Materials to this article. The designation of VSH terms adopted here follows the scheme {\bf MVSH}$_{klm}\equiv\{mag,k,l,m\}$ and {\bf EVSH}$_{klm}\equiv\{ele,k,l,m\}$, with $mag$ standing for magnetic (toroidal, or curl) type, $ele$ for electric (poloidal, divergence) type, $k=0,\;1,\;2$ designating the kind, $l=1,\ldots , L$ being the degree of VSH, and $m=0,\dots, l$ the order of VSH. Kind $k=0$ is reserved for the zonal harmonics only, which have $m=0$. Kind $k=1$ includes harmonics derived from the real parts of the original complex-valued VSH functions. Kind $k=2$ includes harmonics derived from the imaginary parts of the original VSH functions. Consequently, the Fourier parts of these two kinds are mutually orthogonal in the first angular coordinate (longitude). 

For this study, we used the cell-averaging technique within a grid of nearly equal-area cells on the sphere \citep{2019AJ....158..158M, 2021MNRAS.506.5540M} to speed up the computation. This cell-averaging does not significantly affect the fitting coefficients within the domain of low-order VSH functions, which have $>20$ times longer characteristic wavelengths, while it reduces the random noise of the input measurement vectors. The cells of approximately 4.5 by 4.5 squared degrees in size are small enough to preserve the amplitudes of the third-degree VSH fits. The weights of individual proper motions, which are $2\times 2$ matrices, are computed in the traditional way from the formal covariances. The formal uncertainties of the derived average proper motions are obtained from the Maximum Likelihood condition under the assumption, that the distribution of the individual proper motion vectors follows the binormal distribution \citep{2022AJ....164..157M}, which is approximately correct for the CRF sample. The inverse of the the cell-averaged covariance matrices are then used as $2\times 2$ weights in the VSH solution. This is a crucial step, because the cells have widely different formal uncertainties, mostly due to the uneven distribution of sources on the sky. 

In each of the VSH solutions described in this paper, only cells with more than 14 data vectors are used, eliminating a number of cells along the Galactic equator. The entire analysis is done in the Galactic coordinate system $\{l,\, b\}$ to facilitate the interpretation of possible signals such as the Galactic aberration signal in the first-degree electric harmonics \citep{2003A&A...404..743K, 2006AJ....131.1471K}. To this end, the individual proper motions and their covariance matrices are transformed from the original equatorial system (ICRS) to the Galactic coordinate system following standard procedures. The averaging is performed for each cell separately. The VSH fit is obtained by solving the constructed system of linear equations by the weighted least-squares method. The size of the design matrix is $N_{\rm cell}\times N_{\rm VSH}\times 2$. Because of the tri-dimensional system of equations, standard LS routines may not be used unless a large block-diagonal weight matrix is explicitly generated along with the flattened design matrix. In this study, ad hoc software was employed to compute the normal equations and the weighted right-hand parts (which are $N_{\rm cell}\times 2$ matrices). The result is a vector of $N_{\rm VSH}=2 L (2+L)$ coefficients in units of \uasyr. 

This analysis is implemented with a limiting VSH degree of $L=3$. It is chosen from the considerations of the general robustness of the fit. The total number of fitted VSH functions (30) is significantly smaller than the number of cell points, which allows to increase the signal-to-noise ratio (S/N) of the estimated coefficients. The limited degree of the VSH decompositions means that only the large-scale or global patterns of tangential motion can be probed. 

\section*{Data Availability}
The complete catalogue of ML-synthesized and spectroscopically measured redshifts for 1.5 million Gaia CRF3 quasars and active galactic nuclei is available in Supplementary Data 1 and via Zenodo at \url{https://doi.org/10.5281/zenodo.15306518}, \citep{makarov_2025_15306518}.
The Gaia DR3 CRF catalog is available via the {\it VizieR} web service for download and ADQL query as catalog I/355/gcrf3xm, \url{http://vizier.cds.unistra.fr/viz-bin/VizieR-3?-source=I/355/gcrf3xm}. Additional astrometric and photometric information was extracted from the main source Gaia DR3 catalog I/355/gaiadr3, \url{http://vizier.cds.unistra.fr/viz-bin/VizieR-3?-source=I/355/gaiadr3}. The unWISE catalog of MIR-photometry is avaiable as II/363, \url{http://vizier.cds.unistra.fr/viz-bin/VizieR?-source=II/363&-to=3}.
The SDSS quasar catalog DR16 is listed as VII/289/dr16q, \url{http://vizier.cds.unistra.fr/viz-bin/VizieR-3?-source=VII/289/dr16q}. Access to the Quaia catalog of quasar redshifts, used for independent verification of these results, is available at \url{https://doi.org/10.5281/zenodo.10403370}.

\bibliography{manuscript}
\bibliographystyle{aasjournal}

\section{Appendix}
\subsection{Vector spherical harmonic decomposition of astronomical vector fields}
 Astrometric proper motions and position differences between two astrometric catalogs can often be approximated as sets of small three-dimensional vectors tangential to the celestial sphere at the points of their origins. Such sets of vectors, generally called vector fields, contain a sky-correlated component, which depends on the celestial position and is correlated between nearby sources, and a stochastic component without any local correlation or deterministic dependence on celestial position. The sky-correlated component can be of artificial or physical nature. For example, rigid rotation of the coordinate frame by a small amount is an example of a common distortion resulting in positional displacements between two catalogs, which depends on the orientation of the axis of rotation. Vector fields of any complexity can be represented with a sufficient number of vector spherical harmonics (VSH) that are the vector-valued basis functions on the unit sphere. Rigid rotation (as long as it is small) is exactly and completely represented by three specific low-degree terms from this basis. Higher-degree VSH terms represent continuous vector field patterns of increasing complexity. In this section, the mathematical foundation of this technique is described in detail.

\subsubsection{Starting definitions}
\label{sec1.sec}
In standard spherical coordinates $\{r,\theta,\phi\}$, a vector field can be expanded as
\eb
\boldsymbol u (r,\theta,\phi)=\sum_{l=0}^{+\infty}\sum_{m=-l}^{+l}\; u_m^l(r)\boldsymbol{R}_m^l(\theta,\phi)+
v_m^l(r)\boldsymbol{S}_m^l(\theta,\phi)+
w_m^l(r)\boldsymbol{T}_m^l(\theta,\phi)
\label{u.eq}
\ee
with
\begin{eqnarray}
\boldsymbol{R}_m^l(\theta,\phi)&=& Y_m^l(\theta,\phi)\boldsymbol{e}(r,\theta,\phi) \nonumber \\
\boldsymbol{S}_m^l(\theta,\phi)&=& \nabla \, Y_m^l(\theta,\phi) \nonumber \\
\boldsymbol{T}_m^l(\theta,\phi)&=& \nabla \times Y_m^l(\theta,\phi)
\label{rst.eq}
\end{eqnarray}
defined on a unit sphere via the usual normalized scalar spherical harmonics (VSHs) $Y_m^l(\theta,\phi)$, also defined on a unit sphere, where $\boldsymbol{e}(r,\theta,\phi)$ is the unit vector from the center in the direction of point $(r,\theta,\phi)$. The infinite set of vector functions $\boldsymbol{R}$, $\boldsymbol{S}$, and $\boldsymbol{T}$ are called the vectorial spherical harmonics. The normalized scalar spherical harmonics satisfy the differential equation
\eb
\frac{\partial^2 Y_m^l}{\partial\theta^2}+\cot\theta \frac{\partial Y_m^l}{\partial\theta}+\frac{1}{\sin^2\theta}\frac{\partial^2 Y_m^l}{\partial\phi^2}+l(l+1) Y_m^l=0.
\label{eq.eq}
\ee
The three kinds of VSHs are mutually orthogonal on the sphere, which means that their inner dot products integrated over the sphere surface are all equal to 0 for any combinations of degrees $m$ and orders $l$. The VSHs of the same kind are also intrinsically orthogonal with the inner products $\boldsymbol{R}_m^l\cdot \boldsymbol{\bar R}_{m'}^{l'}$ integrating to $\delta_{mm'}\delta_{ll'}$ and both $\boldsymbol{S}_m^l\cdot \boldsymbol{\bar S}_{m'}^{l'}$ and $\boldsymbol{T}_m^l\cdot \boldsymbol{\bar T}_{m'}^{l'}$ integrating to $l(l+1)\delta_{mm'}\delta_{ll'}$, with the horizontal bar denoting complex conjugate. Therefore, the set of VSHs represents a basis of vector-valued functions, which is a valuable property finding numerous applications in mathematical physics, celestial mechanics and hydrodynamics. Another valuable property is that the three kinds are easily transformable to each other by the curl operator $\nabla \times$ \citep{1987GApFD..39..163R}, but this property has no applications in astrometry, to my knowledge.

The VSH expansion in its complete form in Eq. \ref{u.eq} can be used to represent astronomical 3D vector fields within a spherical volume, for example, to describe the velocity field of Galactic stars within a limited volume around the Sun, but I am not aware of such applications in the literature. For classical astrometry, the celestial sphere is modeled as a unit sphere centered on the Solar system barycenter, and celestial sources are points on this sphere defined by unit position vectors. Position offsets, as well as their time derivatives (called proper motions) are well approximated by vectors tangential to the true position of a source at a fixed time $t_0$. Thus, the position unit vector $\boldsymbol{r}(t_0)$ and the proper motion vector $\boldsymbol\mu(t_0)=d  \boldsymbol{r}(t_0)/dt$ are orthogonal. Both vectors, however, are epoch-dependent (time variable). In principle, proper motion vectors for the same source in two astrometric catalogs cannot be directly compared if they refer to different epochs or even different mean positions. The tangential direction of astrometric vectors simplify the VSH decomposition removing completely the radial component $\boldsymbol{R}_m^l$ and rendering the coefficients $v_m^l$ and $w_m^l$ constants rather than functions of radius. We note that the normalization coefficient $1/\sqrt{l(l+1)}$ is inserted in the definition of the vector harmonics in some papers, e.g., \citep{2012A&A...547A..59M}. The $\boldsymbol{S}_m^l$ harmonics are often called spheroidal (or electric, as in this paper), and $\boldsymbol{T}_m^l$ are toroidal (or magnetic) harmonics.

The complex-valued spherical harmonics $Y_m^l(\theta,\phi)$ satisfy the angular part of Laplace's equation in spherical coordinates, which is called the spherical harmonic equation (\ref{eq.eq}). They are defined as \citep{garfken67:math}
\eb
Y_m^l(\theta,\phi)\equiv \sqrt{\frac{2 l+1}{4 \pi}\frac{(l-m)!}{(l+m)!}} \;P_l^m(\cos \theta) e^{im\phi},
\label{y.eq}
\ee
with $l\ge 0$, $m=-l,-l+1,\ldots,+l$, and $P_l^m(x)$ being the associated Legendre polynomials. Spherical harmonics are orthogonal on the unit sphere and normalized to unity with a $\sin\theta$ weight:
\eb
\int_0^{2\pi}\int_0^\pi\;Y_l^m(\theta,\phi) \bar Y_{l'}^{m'}(\theta,\phi)\,\sin \theta = \delta_{mm'}\delta_{ll'}.
\ee
The Condon-Shortley phase $(-1)^m$ is added as a factor in the definition of spherical harmonics in some papers \citep[e.g.,][]{2012A&A...547A..59M}. A brief introduction to scalar spherical harmonics is given in \url{https://mathworld.wolfram.com/SphericalHarmonic.html}.

 \subsubsection{Applications of VSH in astrometry}

Perhaps the first documented application of scalar spherical harmonics (SSHs) in astrometry was proposed by \citet{1966VeARI..17....1B}, who proposed to use a finite set of these basis function for systematic differences in catalog positions or proper motions. As discussed in the preceding Section, these differences are in fact tangential vector fields on the unit sphere, and P. Brosche resolved this difficulty by using a separate and independent set SSHs for the two coordinate components of the tangential vectors. This approach has inherent flaws because it creates an excessive number of fitting functions and is subject to singularities at the celestial poles. Furthermore, such dual representation with independent scalar harmonics is not invariant with respect to rotation of the coordinate system, resulting in different outcomes with different astronomical coordinate systems. These problems are resolved with the VSHs. The idea, however, took some time to get traction in the astrometric community. The initial suggestion by \citet{1990srst.conf..151M} was further developed by \citet{2004ASPC..316..230V} and \citet{2006jsrs.conf...41S}. \citet{2007AJ....134..367M} computed a VSH expansion of the global field of Hipparcos proper motions and demonstrated how the fundamental parameters of the Ogorodnikov-Milne tensor model of Galactic kinematics are related to the fitting coefficients. VSH terms can be directly estimated as independent unknowns describing the sky-correlated and systematic errors of vector fields in global astrometric solutions \citep{2012AJ....144...22M}, a potential that has yet to be realized.

The objective of this memorandum is to provide information about the technical aspects of VSH implementation, so required modifications to the previously described formalism will now be discussed. We note that the spherical coordinate $\theta$ in the equations above is the south pole distance (colatitude), while astronomical coordinates include latitude angles reckoned from a reference midplane (e.g., declination $\delta$ with respect to the equator). The transformation to the equatorial coordinate system is $\phi \rightarrow \alpha$, $\theta \rightarrow \delta +\pi/2$. From Eqs. \ref{rst.eq}, the tangential VSHs can then be defined as
\begin{eqnarray}
\boldsymbol{S}_m^l(\theta,\phi)&=& \left[ \frac{1}{\cos\delta}\frac{\partial Y_m^l(\alpha,\delta)}{\partial\alpha} \boldsymbol\tau_\alpha+ \frac{\partial Y_m^l(\alpha,\delta)}{\partial\delta}\boldsymbol\tau_\delta\right] \nonumber \\
\boldsymbol{T}_m^l(\theta,\phi)&=& \left[ \frac{\partial Y_m^l(\alpha,\delta)}{\partial\delta}\boldsymbol\tau_\alpha- \frac{1}{\cos\delta}\frac{\partial Y_m^l(\alpha,\delta)}{\partial\alpha}\boldsymbol\tau_\delta\right]
\label{def.eq}
\end{eqnarray}
where the unit vectors $\boldsymbol\tau_\alpha$ and $\boldsymbol\tau_\delta$ are the tangential components of the local coordinate triad in the denominations of \citep{2007AJ....134..367M}, also cf. Fig. 1 in \citep{2012A&A...547A..59M}. $\boldsymbol\tau_\alpha$ is directed to the east as seen by the observer on the sky, and $\boldsymbol\tau_\delta$ is directed toward the north pole. In advanced computer languages such as Mathematica, Eqs. \ref{def.eq} are sufficient to express VSHs as trigonometric expansions of the angular coordinates, as well as to compute their values for any point on the unit sphere.

Astrometric vector fields are real-valued vector functions, so the complex presentation in Eqs. \ref{y.eq} and \ref{def.eq} is excessive. We note that all the spherical harmonics of zeroth order ($m=0$) are real-valued, while for $m\neq 0$, $Y_l^{-m}(\theta,\phi)=\bar Y_l^{+m}(\theta,\phi)$ (cf. Eq. \ref{y.eq}). Without a loss of completeness, the complex VSHs with negative orders can be removed and real-valued real and imaginary parts of VSHs with positive orders used. The complete infinite expansion of a real-valued field $\boldsymbol u$ is then
\eb
\begin{split}
\boldsymbol u (\alpha,\delta)= \sum_{l=1}^{+\infty} [ & c_{0l0}{\bf EVSH}_{0l0}(\alpha,\delta)+ d_{0l0}{\bf MVSH}_{0l0}(\alpha,\delta) \\
&+\sum_{k=1}^2 \sum_{m=1}^l\; c_{klm}{\bf EVSH}_{klm}(\alpha,\delta) +
d_{klm}{\bf MVSH}_{klm}(\alpha,\delta) ]
\end{split}
\label{vsh.eq}
\ee
where
\begin{eqnarray}
{\bf EVSH}_{0l0}&=&\boldsymbol S_l^0 \nonumber \\
{\bf MVSH}_{0l0}&=&\boldsymbol T_l^0 \nonumber \\
{\bf EVSH}_{1lm}&=&{\mathfrak{Re}}\left[\boldsymbol S_l^m\right] \nonumber \\
{\bf MVSH}_{1lm}&=&\mathfrak{Re} \left[\boldsymbol T_l^m\right] \nonumber \\
{\bf EVSH}_{2lm}&=&\mathfrak{Im}\left[\boldsymbol S_l^m\right] \nonumber \\
{\bf MVSH}_{2lm}&=&\mathfrak{Im}\left[\boldsymbol T_l^m\right].
\end{eqnarray}
In practical applications, these VSH expansions are truncated at a certain degree $L$. The number of independent and mutually orthogonal harmonics for each degree $l$ is $2(1+2l)$, so that the total number of terms in a truncated expansion is $N=2L(2+L)$. Note that the number of fitting terms quadratically increases with $L$. Using Wolfram Mathematica, analytical expressions can be quickly found for VSHs of practically any degree and order. For example,
\eb
{\bf MVSH}_{296}=\left\{\begin{aligned} &
\frac{3}{1024}\sqrt{\frac{40755}{\pi}}\sin(6 \alpha)\cos^5\delta (65 - 108 \cos(2 \delta) + 51 \cos(4 \delta) )\\ &
\frac{3}{256}\sqrt{\frac{40755}{\pi}}\cos(6 \alpha)\cos^5\delta (39 \sin \delta - 17 \sin(3 \delta) )
\end{aligned}\right\}
\ee
Note that each VSH is a 2-vector composed of the projections onto $\boldsymbol\tau_\alpha$ and $\boldsymbol\tau_\delta$.

\subsubsection{Practical considerations: The choice of coordinate system}
A few issues of practical significance should be taken into account to achieve an accurate and unambiguous implementation of VSHs in analysis of astrometric vector fields. 
The same vector field $\boldsymbol u$ can be rotated between different celestial coordinate systems. The usual choice for astronomers is the equatorial ICRS system $(\alpha,\delta)$, initially based on the mean equator at J2000, the ecliptic system $(\lambda,\beta)$, and the Galactic system $(\ell,b)$, which refers to a nominal fixed Galactic plane and Galactic center direction. The rotation matrices needed for transformations between these three systems of spherical coordinates are given, for example, in \citep[][Vol.1]{1997ESASP1200.....E}. These transformations apply to any vectors, including the source-specific triads $\{\boldsymbol\tau_\alpha,\boldsymbol\tau_\delta,\boldsymbol r\}$. Therefore, a VSH fit in one coordinate system will be different from a fit to the same vector field in another coordinate system. In other words, the fitting coefficients $c_{klm}$, $d_{klm}$ obtained in the Galactic system cannot be directly compared to their counterparts in the equatorial system. The objectives and expected signals for a specific study often dictate the choice of the coordinate system. For example, the secular aberration effect caused by the acceleration of the Solar system barycenter in the Milky Way's potential is expected the be observable as a dipole pattern represented by only three VSHs, viz., $\bf{EVSH}_{010}$, $\bf{EVSH}_{111}$, and $\bf{EVSH}_{211}$ \citep{2006AJ....131.1471K}. They are expected to take the simplest form (with only one significant term) in the Galactic system. Propagation of systematic and sky-correlated errors in the Gaia mission, on the other hand, is more conveniently analyzed in the ecliptic system because these errors are likely to be related to the Gaia scanning law and the position of the Sun. There is a remarkable property that any VSH upon rigid rotation can be represented by a limited series (linear combination) of VSHs of the same degree $l$ \citep[][Sect. 3]{2012A&A...547A..59M}. However, it is more practical and less risky to rotate the input data $\boldsymbol u$ to the selected system.

\subsubsection{Orthogonality and normalization}
\label{ort.sec}
Both the orthogonality and normalization properties of VSHs in the Hilbert space of continuous vector functions, discussed in Section \ref{sec1.sec}, do not hold when these functions are sampled on a discreet data set. This is the situation with astrometric vector fields, which are defined on discrete sets of specific source positions. Thus, if the integrals over unit sphere in the definition of the inner product (and distance between the functions) are replaced with summation over the data points, the products of VSHs are not equal to zero or the number of points $N$. The main reason is that the data points are never distributed uniformly enough on the celestial sphere. Furthermore, the data vectors are of unequal precision, and should be unequally weighted in the least-squares fit. Consequently, expansions \ref{vsh.eq} cannot be computed by simple projection and require full-scale optimization procedures. In the most readily available least-squares solution, the emerging covariance matrix includes nonzero off-diagonal elements, and the diagonal elements (variances) are not equal. The singular values of the design matrix are dispersed within a certain range resulting in a structural weakness of the solution and loss of robustness \citep{2021AJ....161..289M}, which depends, among other parameters, on the number of fitting VSHs.

\subsubsection{Ordering scheme}
\label{ord.sec}
Expansions \ref{vsh.eq} can be determined for any finite subset of the infinite manifold of VSHs, but it is practical and customary to use a complete set limited to a certain degree $L$. With a finite number of data points on the celestial sphere, the low-degree terms are usually more accurately determined, and any detected signal is more likely to be real. Very roughly, the upper limit on the number of fitting functions is set by the number of data vectors $N$. In practice, a much smaller number of terms should be involved because high-degree harmonics become ill-conditioned on discrete data sets. Some recommendations about selecting $L$ are given in \citep{2021AJ....161..289M}. 

It is convenient to put Eqs. \ref{vsh.eq} into a matrix form for a least-squares adjustment. The problem can be written as
\eb
\boldsymbol{A}\;\boldsymbol{x}=\boldsymbol{y},
\label{a.eq}
\ee
where $\boldsymbol{A}$ is the design matrix with $2L(2+L)$ columns, $\boldsymbol{x}$ is the vector of unknown coefficients $c_{klm}$ and $d_{klm}$, and $\boldsymbol{y}$ is the vector of observations. The order of VSH functions in the columns of $\boldsymbol{A}$, which is identical to the order of the unknown coefficients, can be arbitrary, in principle. It is important, however, to avoid confusion with the adopted ordering scheme in the interpretation of the results. Here, as guidelines, I describe the scheme implemented in my Mathematica software and used in this paper. Batches of columns are ordered by increasing degree $l=1,2,\ldots,L$. Each batch includes all orders for a given $l$. Within each batch, the harmonics are divided into groups, one group for each $m$ in increasing order, $m=1,\ldots, l$. Within each group, all the magnetic terms are listed first followed by the electric terms. Thus, for $l=2$, the corresponding batch includes the functions (\bf{MVSH}$_{020}$,  \bf{MVSH}$_{121}$, \bf{MVSH}$_{221}$, \bf{EVSH}$_{020}$,  \bf{EVSH}$_{121}$, \bf{EVSH}$_{221}$, \bf{MVSH}$_{122}$, \bf{MVSH}$_{222}$, \bf{EVSH}$_{122}$, \bf{EVSH}$_{222}$\rm)
, 10 in total. The specific VSH functions up to degree 4 in the adopted order are listed in Table \ref{vshlist.tab}.

The dimensions of $\boldsymbol{A}$ are $\{N,2L(2+L),2\}$, i.e., it is a 3D matrix. Likewise, the dimensions of $\boldsymbol{y}$ are $\{N,2\}$, because each observation is a 2-vector. Therefore, standard library functions for solving least-squares problems would not work. This is not critical, because the LS solution is simply
\eb
\widehat{\boldsymbol{x}}=(\boldsymbol{A}^T \boldsymbol{A})^{-1}\,
(\boldsymbol{A}^T\,\boldsymbol{y}),
\label{ls.eq}
\ee 
where in calculating the covariance matrix $\boldsymbol{C}=(\boldsymbol{A}^T \boldsymbol{A})^{-1}$ and the projection $\boldsymbol{A}^T\,\boldsymbol{y}$ we replace the scalar multiplication with the inner product of the corresponding 2-vectors. This requires a little more programming toil but is not taxing in terms of the computing time. Alternatively, the $\boldsymbol \tau_{\alpha}$ components of the data vectors in the right-hand part can be separated from the $\boldsymbol \tau_{\delta}$ components and these parts stacked forming a longer vector with $2 N$ elements. A similar transformation results in a $2 N\times 2L(2+L)$ rectangular array for a design matrix. In principle, these alternative techniques should produce identical results for unweighted solutions but weighting schemes may be different, as discussed in the next paragraph.

\newpage
\startlongtable
\begin{deluxetable*}{L|L|C}
\tablecaption{VSH functions to degree 4.}
 \label{vshlist.tab}
\tablehead{
\colhead{j} & \colhead{\text{name}} & \colhead{\text{function}}
}
\startdata
 1 & \{\text{mag},0,1,0\} & \left\{-\frac{1}{2} \sqrt{\frac{3}{\pi }} \cos (b),0\right\}
   \\
 2 & \{\text{mag},1,1,1\} & \left\{\frac{1}{2} \sqrt{\frac{3}{2 \pi }} \sin (b) \cos
   (l),-\frac{1}{2} \sqrt{\frac{3}{2 \pi }} \sin (l)\right\} \\
 3 & \{\text{mag},2,1,1\} & \left\{\frac{1}{2} \sqrt{\frac{3}{2 \pi }} \sin (b) \sin
   (l),\frac{1}{2} \sqrt{\frac{3}{2 \pi }} \cos (l)\right\} \\
 4 & \{\text{ele},0,1,0\} & \left\{0,-\frac{1}{2} \sqrt{\frac{3}{\pi }} \cos (b)\right\}
   \\
 5 & \{\text{ele},1,1,1\} & \left\{\frac{1}{2} \sqrt{\frac{3}{2 \pi }} \sin
   (l),\frac{1}{2} \sqrt{\frac{3}{2 \pi }} \sin (b) \cos (l)\right\} \\
 6 & \{\text{ele},2,1,1\} & \left\{-\frac{1}{2} \sqrt{\frac{3}{2 \pi }} \cos
   (l),\frac{1}{2} \sqrt{\frac{3}{2 \pi }} \sin (b) \sin (l)\right\} \\
 7 & \{\text{mag},0,2,0\} & \left\{\frac{3}{2} \sqrt{\frac{5}{\pi }} \sin (b) \cos
   (b),0\right\} \\
 8 & \{\text{mag},1,2,1\} & \left\{\frac{1}{2} \sqrt{\frac{15}{2 \pi }} \cos (2 b) \cos
   (l),\frac{1}{2} \sqrt{\frac{15}{2 \pi }} \sin (b) \sin (l)\right\} \\
 9 & \{\text{mag},2,2,1\} & \left\{\frac{1}{2} \sqrt{\frac{15}{2 \pi }} \cos (2 b) \sin
   (l),-\frac{1}{2} \sqrt{\frac{15}{2 \pi }} \sin (b) \cos (l)\right\} \\
 10 & \{\text{ele},0,2,0\} & \left\{0,\frac{3}{2} \sqrt{\frac{5}{\pi }} \sin (b) \cos
   (b)\right\} \\
 11 & \{\text{ele},1,2,1\} & \left\{-\frac{1}{2} \sqrt{\frac{15}{2 \pi }} \sin (b) \sin
   (l),\frac{1}{2} \sqrt{\frac{15}{2 \pi }} \cos (2 b) \cos (l)\right\} \\
 12 & \{\text{ele},2,2,1\} & \left\{\frac{1}{2} \sqrt{\frac{15}{2 \pi }} \sin (b) \cos
   (l),\frac{1}{2} \sqrt{\frac{15}{2 \pi }} \cos (2 b) \sin (l)\right\} \\
 13 & \{\text{mag},1,2,2\} & \left\{-\frac{1}{4} \sqrt{\frac{15}{2 \pi }} \sin (2 b)
   \cos (2 l),\sqrt{\frac{15}{2 \pi }} \cos (b) \sin (l) \cos (l)\right\} \\
 14 & \{\text{mag},2,2,2\} & \left\{-\sqrt{\frac{15}{2 \pi }} \sin (b) \cos (b) \sin (l)
   \cos (l),-\frac{1}{2} \sqrt{\frac{15}{2 \pi }} \cos (b) \cos (2 l)\right\} \\
 15 & \{\text{ele},1,2,2\} & \left\{-\sqrt{\frac{15}{2 \pi }} \cos (b) \sin (l) \cos
   (l),-\frac{1}{4} \sqrt{\frac{15}{2 \pi }} \sin (2 b) \cos (2 l)\right\} \\
 16 & \{\text{ele},2,2,2\} & \left\{\frac{1}{2} \sqrt{\frac{15}{2 \pi }} \cos (b) \cos
   (2 l),-\sqrt{\frac{15}{2 \pi }} \sin (b) \cos (b) \sin (l) \cos (l)\right\} \\
 17 & \{\text{mag},0,3,0\} & \left\{\frac{3}{8} \sqrt{\frac{7}{\pi }} \cos (b) (5 \cos
   (2 b)-3),0\right\} \\
 18 & \{\text{mag},1,3,1\} & \left\{\frac{1}{32} \sqrt{\frac{21}{\pi }} (\sin (b)-15
   \sin (3 b)) \cos (l),\frac{1}{16} \sqrt{\frac{21}{\pi }} (5 \cos (2 b)-3) \sin
   (l)\right\} \\
 19 & \{\text{mag},2,3,1\} & \left\{\frac{1}{32} \sqrt{\frac{21}{\pi }} (\sin (b)-15
   \sin (3 b)) \sin (l),\frac{1}{8} \sqrt{\frac{21}{\pi }} \left(5 \sin ^2(b)-1\right)
   \cos (l)\right\} \\
 20 & \{\text{ele},0,3,0\} & \left\{0,\frac{3}{8} \sqrt{\frac{7}{\pi }} \cos (b) (5 \cos
   (2 b)-3)\right\} \\
 21 & \{\text{ele},1,3,1\} & \left\{\frac{1}{8} \sqrt{\frac{21}{\pi }} \left(5 \sin
   ^2(b)-1\right) \sin (l),\frac{1}{32} \sqrt{\frac{21}{\pi }} (\sin (b)-15 \sin (3 b))
   \cos (l)\right\} \\
 22 & \{\text{ele},2,3,1\} & \left\{\frac{1}{16} \sqrt{\frac{21}{\pi }} (5 \cos (2 b)-3)
   \cos (l),\frac{1}{32} \sqrt{\frac{21}{\pi }} (\sin (b)-15 \sin (3 b)) \sin
   (l)\right\} \\
 23 & \{\text{mag},1,3,2\} & \left\{-\frac{1}{16} \sqrt{\frac{105}{2 \pi }} (\cos (b)+3
   \cos (3 b)) \cos (2 l),-\sqrt{\frac{105}{2 \pi }} \sin (b) \cos (b) \sin (l) \cos
   (l)\right\} \\
 24 & \{\text{mag},2,3,2\} & \left\{-\frac{1}{16} \sqrt{\frac{105}{2 \pi }} (\cos (b)+3
   \cos (3 b)) \sin (2 l),\frac{1}{4} \sqrt{\frac{105}{2 \pi }} \sin (2 b) \cos (2
   l)\right\} \\
 25 & \{\text{ele},1,3,2\} & \left\{\sqrt{\frac{105}{2 \pi }} \sin (b) \cos (b) \sin (l)
   \cos (l),-\frac{1}{16} \sqrt{\frac{105}{2 \pi }} (\cos (b)+3 \cos (3 b)) \cos (2
   l)\right\} \\
 26 & \{\text{ele},2,3,2\} & \left\{-\frac{1}{4} \sqrt{\frac{105}{2 \pi }} \sin (2 b)
   \cos (2 l),-\frac{1}{16} \sqrt{\frac{105}{2 \pi }} (\cos (b)+3 \cos (3 b)) \sin (2
   l)\right\} \\
 27 & \{\text{mag},1,3,3\} & \left\{\frac{3}{8} \sqrt{\frac{35}{\pi }} \sin (b) \cos
   ^2(b) \cos (3 l),-\frac{3}{8} \sqrt{\frac{35}{\pi }} \cos ^2(b) \sin (3 l)\right\} \\
 28 & \{\text{mag},2,3,3\} & \left\{\frac{3}{8} \sqrt{\frac{35}{\pi }} \sin (b) \cos
   ^2(b) \sin (3 l),\frac{3}{8} \sqrt{\frac{35}{\pi }} \cos ^2(b) \cos (3 l)\right\} \\
 29 & \{\text{ele},1,3,3\} & \left\{\frac{3}{8} \sqrt{\frac{35}{\pi }} \cos ^2(b) \sin
   (3 l),\frac{3}{8} \sqrt{\frac{35}{\pi }} \sin (b) \cos ^2(b) \cos (3 l)\right\} \\
 30 & \{\text{ele},2,3,3\} & \left\{-\frac{3}{8} \sqrt{\frac{35}{\pi }} \cos ^2(b) \cos
   (3 l),\frac{3}{8} \sqrt{\frac{35}{\pi }} \sin (b) \cos ^2(b) \sin (3 l)\right\} \\
 31 & \{\text{mag},0,4,0\} & \left\{\frac{15 \sin (b) \left(7 \sin ^2(b)-3\right) \cos
   (b)}{4 \sqrt{\pi }},0\right\} \\
 32 & \{\text{mag},1,4,1\} & \left\{\frac{3}{16} \sqrt{\frac{5}{\pi }} (\cos (2 b)-7
   \cos (4 b)) \cos (l),\frac{3}{32} \sqrt{\frac{5}{\pi }} (9 \sin (b)-7 \sin (3 b))
   \sin (l)\right\} \\
 33 & \{\text{mag},2,4,1\} & \left\{\frac{3}{16} \sqrt{\frac{5}{\pi }} (\cos (2 b)-7
   \cos (4 b)) \sin (l),\frac{3}{8} \sqrt{\frac{5}{\pi }} \sin (b) \left(3-7 \sin
   ^2(b)\right) \cos (l)\right\} \\
 34 & \{\text{ele},0,4,0\} & \left\{0,\frac{15 \sin (b) \left(7 \sin ^2(b)-3\right) \cos
   (b)}{4 \sqrt{\pi }}\right\} \\
 35 & \{\text{ele},1,4,1\} & \left\{\frac{3}{8} \sqrt{\frac{5}{\pi }} \sin (b) \left(3-7
   \sin ^2(b)\right) \sin (l),\frac{3}{16} \sqrt{\frac{5}{\pi }} (\cos (2 b)-7 \cos (4
   b)) \cos (l)\right\} \\
 36 & \{\text{ele},2,4,1\} & \left\{\frac{3}{32} \sqrt{\frac{5}{\pi }} (9 \sin (b)-7
   \sin (3 b)) \cos (l),\frac{3}{16} \sqrt{\frac{5}{\pi }} (\cos (2 b)-7 \cos (4 b))
   \sin (l)\right\} \\
 37 & \{\text{mag},1,4,2\} & \left\{\frac{3}{16} \sqrt{\frac{5}{2 \pi }} (2 \sin (2 b)+7
   \sin (4 b)) \cos (2 l),\frac{3}{16} \sqrt{\frac{5}{2 \pi }} (3 \cos (b)-7 \cos (3 b))
   \sin (2 l)\right\} \\
 38 & \{\text{mag},2,4,2\} & \left\{\frac{3}{16} \sqrt{\frac{5}{2 \pi }} (2 \sin (2 b)+7
   \sin (4 b)) \sin (2 l),\frac{3}{16} \sqrt{\frac{5}{2 \pi }} (7 \cos (3 b)-3 \cos (b))
   \cos (2 l)\right\} \\
 39 & \{\text{ele},1,4,2\} & \left\{\frac{3}{16} \sqrt{\frac{5}{2 \pi }} (7 \cos (3 b)-3
   \cos (b)) \sin (2 l),\frac{3}{16} \sqrt{\frac{5}{2 \pi }} (2 \sin (2 b)+7 \sin (4 b))
   \cos (2 l)\right\} \\
 40 & \{\text{ele},2,4,2\} & \left\{\frac{3}{16} \sqrt{\frac{5}{2 \pi }} (3 \cos (b)-7
   \cos (3 b)) \cos (2 l),\frac{3}{16} \sqrt{\frac{5}{2 \pi }} (2 \sin (2 b)+7 \sin (4
   b)) \sin (2 l)\right\} \\
 41 & \{\text{mag},1,4,3\} & \left\{\frac{3}{8} \sqrt{\frac{35}{\pi }} \cos ^2(b) (2
   \cos (2 b)-1) \cos (3 l),\frac{9}{8} \sqrt{\frac{35}{\pi }} \sin (b) \cos ^2(b) \sin
   (3 l)\right\} \\
 42 & \{\text{mag},2,4,3\} & \left\{\frac{3}{8} \sqrt{\frac{35}{\pi }} \cos ^2(b) (2
   \cos (2 b)-1) \sin (3 l),-\frac{9}{8} \sqrt{\frac{35}{\pi }} \sin (b) \cos ^2(b) \cos
   (3 l)\right\} \\
 43 & \{\text{ele},1,4,3\} & \left\{-\frac{9}{8} \sqrt{\frac{35}{\pi }} \sin (b) \cos
   ^2(b) \sin (3 l),\frac{3}{8} \sqrt{\frac{35}{\pi }} \cos ^2(b) (2 \cos (2 b)-1) \cos
   (3 l)\right\} \\
 44 & \{\text{ele},2,4,3\} & \left\{\frac{9}{8} \sqrt{\frac{35}{\pi }} \sin (b) \cos
   ^2(b) \cos (3 l),\frac{3}{8} \sqrt{\frac{35}{\pi }} \cos ^2(b) (2 \cos (2 b)-1) \sin
   (3 l)\right\} \\
 45 & \{\text{mag},1,4,4\} & \left\{-\frac{3}{4} \sqrt{\frac{35}{2 \pi }} \sin (b) \cos
   ^3(b) \cos (4 l),\frac{3}{4} \sqrt{\frac{35}{2 \pi }} \cos ^3(b) \sin (4 l)\right\}
   \\
 46 & \{\text{mag},2,4,4\} & \left\{-\frac{3}{4} \sqrt{\frac{35}{2 \pi }} \sin (b) \cos
   ^3(b) \sin (4 l),-\frac{3}{4} \sqrt{\frac{35}{2 \pi }} \cos ^3(b) \cos (4 l)\right\}
   \\
 47 & \{\text{ele},1,4,4\} & \left\{-\frac{3}{4} \sqrt{\frac{35}{2 \pi }} \cos ^3(b)
   \sin (4 l),-\frac{3}{4} \sqrt{\frac{35}{2 \pi }} \sin (b) \cos ^3(b) \cos (4
   l)\right\} \\
 48 & \{\text{ele},2,4,4\} & \left\{\frac{3}{4} \sqrt{\frac{35}{2 \pi }} \cos ^3(b) \cos
   (4 l),-\frac{3}{4} \sqrt{\frac{35}{2 \pi }} \sin (b) \cos ^3(b) \sin (4 l)\right\} 
\enddata
\end{deluxetable*}

\subsubsection{Weighting schemes}
\label{wei.sec}

In regular least-squares solutions with uncorrelated right-hand parts, the optimal weight $\nu_n$, $n=1,\ldots,N$, is inversely proportional to the standard deviation of the corresponding observation. The required optimal weighting is achieved by multiplying each row of the design matrix $\boldsymbol{A}$ and each element of the right-hand vector $\boldsymbol{y}$ by $1/\sigma_n$, i.e., 
\begin{eqnarray}
\boldsymbol{A}' &=& {\rm diag}(\boldsymbol{\nu})\,\boldsymbol{A}
\nonumber \\
\boldsymbol{y}' &=& {\rm diag}(\boldsymbol{\nu})\,\boldsymbol{y}.
\end{eqnarray}
Separating and stacking the two components of the observed and fitting vectors, described in the previous Section, lends itself to a natural weighting scheme where the standard deviations $\sigma_n$ are the formal errors of the corresponding components. For example, if a VSH fit is sought to a proper motion field, the weights are equal to the reciprocal formal errors $\sigma_{\mu\alpha *}$ and $\sigma_{\mu\delta}$. This weighting scheme has a weakness, however. Astrometric vectors (position offsets, proper motions) are bivariate statistics, so even when the data vectors can be assumed to be statistically independent between individual sources, the components of data vectors for a given source are not. This fact is captured by the $2\times 2$ covariance matrix of the corresponding parameters, which is available in the modern major catalogs (Hipparcos, Gaia, ICRF). A more accurate approach is discussed in the following.

With the separated components approach, we have $2 N$ condition equations, where the first block of $N$ equations represents the conditions for one component, and the second block of $N$ equations for the other component. In the generalized weighted least-squares method, the solution is written as
\eb
\widehat{\boldsymbol{x}}=(\boldsymbol{A}^T \,\boldsymbol{W}\, \boldsymbol{A})^{-1}\,
(\boldsymbol{A}^T\,\boldsymbol{W}\,\boldsymbol{y}),
\label{w.eq}
\ee 
where $\boldsymbol{W}$ is a symmetric weight matrix. Ignoring the bivariate nature of the observed vectors results in a diagonal weight matrix. But with the available correlations included, the weight matrix should have nonzero elements outside the diagonal. The generating matrix $\boldsymbol W^{-1}$ is obviously a $2 N\times 2 N$ matrix with the variances in the main diagonal and covariances in two symmetric diagonals separated from the main by $N$ columns. The inverse of such striped matrices is known to be also a striped matrix with as many nonzero diagonals as can fit within preserving the same separation \citep[][Theorem 1]{SMOLARSKI2006416}. Obviously, only two symmetric sub-diagonals can be present in the inverse in our case, so $\boldsymbol{W}$ has the same striped structure as $\boldsymbol W^{-1}$. If $c_1$ and $c_2$ are the diagonal elements of the formal covariance matrix $\boldsymbol c$ for a given observed vector $y_n$, and $c_3$ is the off-diagonal element, then, obviously, they are distributed in the  generating matrix according to $\boldsymbol W^{-1}\{n,n\}=c_1$, $\boldsymbol W^{-1}\{n+N,n+N\}=c_2$, $\boldsymbol W^{-1}\{n,n+N\}=\boldsymbol W^{-1}\{n+N,n\}=c_3$. It is easy to see that the corresponding elements in the weight matrix are $\boldsymbol W\{n,n\}=c'_1$, $\boldsymbol W\{n+N,n+N\}=c'_2$, $\boldsymbol W\{n,n+N\}=\boldsymbol W\{n+N,n\}=c'_3$, where $\boldsymbol c'=\boldsymbol c^{-1}$. Thus, a more accurate weighting scheme is to compute the $2\times 2$ covariance matrices for each observed vector, inverse them, and introduce the resulting weight matrices directly in Eqs. \ref{w.eq}. This technique requires some careful manipulation of the original design matrix and the right-hand side vector $\boldsymbol{y}$, but it is computationally fairly efficient if the weight matrix is split as $\boldsymbol{W}= \boldsymbol{V}^T\,\boldsymbol{V}$ in Eq. \ref{w.eq}. Both $\boldsymbol{A}$ and $\boldsymbol{y}$ can then be pre-multiplied with $\boldsymbol{V}$ and standard, computer efficient routines be used to solve the LS problem. The three diagonals of $\boldsymbol{V}$ are populated with the elements of individual matrices $\boldsymbol{c}_n^{-1/2}$, which is the {\it matrix} square root of the individual covariance of $n$-th observation.

The described weighting scheme is optimal when errors of observation are normally distributed with the formal variances and covariances as given in the source catalogs. In real applications, observational errors are not confined to a Gaussian distribution, with a fraction of data points having much greater deviations that would be quite improbable with a formal normal distribution. These model outliers can strongly perturb a LS solution or even make it completely misleading. In some situations, we may not know {\it a priori}, which data points are such statistical outliers. Therefore, an empirical weighting scheme was proposed in \citep{2021AJ....161..289M}, which does not involve the formal uncertainties at all. Very briefly, the unweighted LS problem (\ref{ls.eq}) is solved on all available data points. The residual vectors are computed as
\eb 
\boldsymbol \rho=\boldsymbol y - \boldsymbol A\,\widehat{\boldsymbol x},
\ee 
and their norms $\rho_n=||\boldsymbol \rho_n ||$. The next iteration is solving a weighted LS problem with weights, which are equal to 1 for data points with $\rho_n\le\rho_{\rm med}$ and $\rho_n/\rho_{\rm med}$ for $\rho_n>\rho_{\rm med}$, where $\rho_{\rm med}$ is the sample median of $\rho$. This method may require a few iteration to converge to a stable solution with updates in the VSH coefficients below a given tolerance. Technically, this weighting scheme is most readily implemented with the vectorial organization of condition equations described in the previous Section, at the cost of somewhat slower computation.

\subsubsection{Relation to infinitesimal rotations of coordinate systems}
\label{rot.sec}

Consider a right-handed coordinate triad $\{\boldsymbol X_1,
\boldsymbol X_2,\boldsymbol X_3\}$ and another triad, $\{\boldsymbol X'_1,
\boldsymbol X'_2,\boldsymbol X'_3\}$, which differs by a rigid rotation. Without a loss of generality, the transformation between these two coordinate systems can be represented by a sequence of right-handed Euler rotations around axes $\boldsymbol X_3$,
$\boldsymbol X_2$, and $\boldsymbol X_3$. Each elementary rotation by angle $z_i$, $i=1,2,3$, is represented by a rotation matrix $\boldsymbol R_i$:
\begin{eqnarray}
\boldsymbol R_1 & = & \left[\begin{array}{ccc}
1 & 0 & 0 \\
0 & \cos{z_1} & -\sin{z_1} \\
0 & \sin{z_1} & \cos{z_1}
\end{array} \right] \nonumber \\
\boldsymbol R_2 & = & \left[\begin{array}{ccc}
\cos{z_2} & 0 & \sin{z_2} \\
0 & 1 & 0 \\
-\sin{z_2} & 0 & \cos{z_2}
\end{array} \right] \nonumber \\
\boldsymbol R_3 & = & \left[\begin{array}{ccc}
\cos{z_3} & -\sin{z_3} & 0 \\
\sin{z_3} & \cos{z_3} & 0 \\
0 & 0 & 1
\end{array} \right] 
\end{eqnarray}
The rotational transformation for any vector $\boldsymbol r$ is then
\eb 
\boldsymbol r=\boldsymbol R_{321}\,\boldsymbol r'=\boldsymbol R_{1}\,\boldsymbol R_{2}\,\boldsymbol R_{3}\,\boldsymbol r'.
\ee 
If the rotation angles $z_i$ are infinitesimally small, we can replace $\sin z_i$ with $z_i$ and $\cos z_i$ with $1$. The resulting rotation matrix in this approximation is
\eb 
\boldsymbol R_{321}  =  \left[\begin{array}{ccc}
1 & -z_3 & z_2 \\
z_3 & 1 & -z_1 \\
-z_2 & z_1 & 1
\end{array} \right]
\ee 
if we keep only terms to $O(z_i)$. A local coordinate triad $\{\boldsymbol r, \boldsymbol \tau_\alpha,  \boldsymbol \tau_\delta\}$ at angular coordinates $(\alpha, \delta)$ obtains from the celestial triad $\{\boldsymbol X_1,
\boldsymbol X_2,\boldsymbol X_3\}$ by rotation $\boldsymbol R_{3}(-\alpha)\,\boldsymbol R_{2}(\delta)$, so that the celestial coordinates of the local radial and tangential vectors are
\eb 
\boldsymbol r  =  \left[\begin{array}{c}
\cos{\alpha}\cos{\delta} \\ \sin{\alpha}\cos{\delta} \\ \sin{\delta} 
\end{array} \right], \quad\boldsymbol \tau_\alpha  =  \left[\begin{array}{c}
-\sin{\alpha} \\ \cos{\alpha} \\ 0 
\end{array} \right], \quad
\boldsymbol \tau_\delta  =  \left[\begin{array}{c}
-\cos{\alpha}\sin{\delta} \\ -\sin{\alpha}\sin{\delta} \\ \cos{\delta}
\end{array} \right]
\label{tau.eq}
\ee 
The offset caused by this rotation in the position vector is
\eb 
\boldsymbol r' - \boldsymbol r = (\boldsymbol R^T_{321}-\boldsymbol I)
\,\boldsymbol r = \left[\begin{array}{c}
z_3\sin{\alpha}\cos{\delta}-z_2\,\sin{\delta} \\ -z_3\cos{\alpha}\cos{\delta}+z_1\,\sin{\delta} \\ 
z_2\cos{\alpha}\cos{\delta}-z_1\,\sin{\alpha}\cos{\delta}
\end{array} \right]
\ee 
Projecting this vector onto the tangential plane spanned by vectors $\boldsymbol \tau_\alpha$ and $\boldsymbol \tau_\delta$ obtains a tangential 2-vector
\eb 
\boldsymbol r_\tau = \left[\begin{array}{c}
-z_3\cos{\delta}+(z_1\,\cos{\alpha}+z_2\,\sin{\alpha})\,\sin{\delta} \\ z_2\cos{\alpha}-z_1\,\sin{\alpha} 
\end{array} \right]
\label{rho.eq}
\ee 
If we compare this result with the explicit trigonometric expressions for the 1st-degree magnetic VSHs, {\bf MVSH}$_{010}=\left[-\frac{1}{2}\sqrt{\frac{3}{\pi}}\cos{\delta},0\right]^T$, {\bf MVSH}$_{111}=\left[\frac{1}{2}\sqrt{\frac{3}{2 \pi}}\cos{\alpha}\sin{\delta},-\frac{1}{2}\sqrt{\frac{3}{2 \pi}}\sin{\alpha} \right]^T$, {\bf MVSH}$_{211}=\left[\frac{1}{2}\sqrt{\frac{3}{2 \pi}}\sin{\alpha}\sin{\delta},\frac{1}{2}\sqrt{\frac{3}{2 \pi}}\cos{\alpha} \right]^T$, we establish the asymptotic equivalence (within a normalization multiplier) of the Euler rotation angles and the corresponding coefficients of the magnetic VSHs, viz., $z_1=2\sqrt{\frac{2 \pi}{3}}\,d_{111}$, $z_2=2\sqrt{\frac{2 \pi}{3}}\,d_{211}$, and $z_3=2\sqrt{\frac{\pi}{3}}\,d_{010}$. Note that this approximation can only be used in practical applications for small rotation angles.

This straightforward correspondence between VSH terms and infinitesimal rigid rotation is important for interpreting position differences between two catalogs, as well as small spins between two proper motion fields, as in this paper. Whenever we compare two reference frames defined by two astrometric catalogs, we would like to know their net relative rotation. The often used approach is to solve the system of linear equations (\ref{rho.eq}) for the entire collection of tangential offset vectors for the three Euler angles $z_1,z_2,z_3$ by a weighted LS method, for example. This is in fact equivalent to performing a truncated VSH decomposition limited to the three magnetic harmonics of 1st degree, which is often done in the published literature. Technically, it is not much simpler than to compute a VSH decomposition that is complete to a certain higher degree. There is a possible loss of accuracy, however. As discussed in Section \ref{ort.sec}, the discretized VSHs are not mutually orthogonal. The covariance matrices of the VSH coefficients are therefore not diagonal, with some nonzero elements outside of the diagonal. When the decomposition is artificially truncated, the intrinsic covariances, which are especially significant between the terms of the same degree, result in propagation of additional error into the fit from adjacent terms. The risk of contamination (discussed as ``harmonic leak" by \citet{2007AJ....134..367M}) is especially high for a rigid rotation determination if there are genuine signals in the electric or magnetic harmonics of low degree that are not accounted for. It is safer to perform a full VSH decomposition to a higher degree and then extract the three magnetic coefficients as the fitted rotation angles.

\subsection{Additional verification tests}
\subsubsection{Complete re-analysis using Quaia-CRF intersection}
The ML-predicted redshifts for 1.6 million Gaia-CRF sources are sufficiently accurate to represent the population of optical quasars in the coarse bins 1--2, 2--3, and $>3$ used in this study. To verify this statement, the entire analysis cycle was repeated using an independent source of redshifts. These redshift values were extracted from the Quaia catalog of 1.3 million confirmed Gaia quasar candidates \citep{2024ApJ...964...69S}. The authors estimate the performance of their regression redshifts in terms of the rate of relative errors $\Delta z/(1+z)$ greater than 0.2, which is at 6\% for the brighter sources with $G<20$ mag. The rate of outliers for the ML-predicted redshifts in this study is somewhat higher but close to this performance. \citep{2024ApJ...964...69S} used a different set of principles for their estimation of redshifts. Therefore, their data provide a good verification test for the results presented here. I cross-matched Quaia with the Gaia-CRF DR3 catalog using the source identifications and obtained $1,125,890$ common sources. This is significantly less than the 1.6 million synthetic redshifts generated in this study, but still enough to reveal the largest signals in the proper motion field. After discarding sources with $z_{\rm Quaia}<1$, the sample dwindles to $838,966$ sources. 

The sequence of data processing steps described in Section \ref{vsh.sec} was applied to this sample after adding the proper motion correlation coefficients from the main Gaia DR3 catalog, which are omitted in Quaia. The sample was divided into 1739 small cells on the celestial sphere. For each cell with more than 10 sources, the weighted mean proper motion vector and the corresponding covariance matrix of the mean were computed. These averaged data vectors were used in a 3D weighted least-squares adjustment of 30 VSH fields up to degree 3. The standard errors of the solution vector were computed from the diagonal of the (weighted) covariance matrix, and the relative significance score $|a_j|/\sigma_{a_j}$ computed for each VSH function (Section \ref{vsh.sec}). The results are given in Table \ref{qvsh.tab} in the same forms as in Table \ref{vsh.tab} for the CRF sample with my own redshifts, for the same three subdivisions by redshift.

We find the results with two independent sets of redshifts on non-identical collections of quasars to be generally consistent within 1--2 estimated formal errors of each VSH coefficient. Larger values seem to emerge for some of the basis functions in the $z>3$ bin, but these estimates are not statistically robust because of the relatively small sample ($24,258$ sources). The most significant signals detected in this paper are confirmed. Comparing Tables \ref{qvsh.tab} and \ref{vsh.tab}, the largest differences are seen in the first three magnetic harmonics for the 2--3 bin where the amplitude of the \{mag,0,1,0\} term became smaller, while the other two magnetic terms became even larger and more discrepant with the 1--2 and the general samples. Thus, the VSH analysis based on the Quaia/CRF sample of quasars supports the main conclusion that we find significant differences in the general spin of distant quasar populations at different cosmological epochs. Unlike the first-degree magnetic harmonics, the electric dipole \{ele,1,1,1\}, which carries the secular acceleration signal, is practically identical to the previous estimates in the most significant redshift bins. The amplitude of this non-cosmological signal from the Quaia/CRF sample is $-5.52\pm0.40$ $\mu$as yr$^{-1}$.

\subsubsection{Alternative methods of ML redshift classification and regression}
The sensitivity of the main results obtained in this study to the particulars of redshift synthesis has been further assessed with additional experiments using different ML regression and classification methods. The results discussed in the main body of this paper are based on the Neural Network (NW) Machine-Learning method of regression. I have reproduced the synthesis of 1.6 million redshifts using the same training set of 0.277 million SDSS-measured values and the methods Linear Regression and Nearest Neighbor (NN). The latter was found to produce similar performance to the Neural Network method. The measures of performance include the 0.16 and 0.84 quantiles of the $z_{\rm obs}-z_{\rm pre}$ errors (cf. Fig. \ref{z-z.fig}) and, more importantly for this study, the percentage of sources that migrated to another sub-sample when the output is divided into $z$-bins 1--2, 2--3, and $>3$. The median curve, which shows the systematic bias of $z_{\rm pre}$ as a function of $z$, is somewhat different from the NW analog, but shows the same main features, sagging at 1.4 and 2.6 and peaking at 2.2. The uncertainty corridor also widens toward higher redshifts at $z>1.4$. We find that 204,163 SDSS sources (73.6\%) remain in the correct $z$-bin with their predicted values, while 13.6\% leak into the lower adjacent bin, and 12.1\% migrate to the higher adjacent bin. This partial mixing of sources between adjacent $z$-bins does not undermine the main conclusions, however. The differential signal in VSH coefficients that we detect here can only be smoothed out by this effect. Indeed, if we perform complete mixing by random permutation of sources, the VSH values should be identical within the standard variance between the bins.

Machine Learning classification provides an interesting alternative to the regression methods, because it is often based on conceptually different principles. Instead of trying to quantify the redshifts of 1.6 million CRF sources, we can directly classify them into the fixed $z$-bins. The choice of methods for ML-classification is more diverse, but only some of them are suitable for numeric data. Neural Network and Nearest Neighbors are also among the available classification methods. I have reproduced the entire duty cycle of this study using a different classification method called Gradient Boosted Trees. In terms of the adjacent bin mixing, this classification achieves marginally better performance, with 75.5\% of the training sample remaining in the correct bin, 12.2\% moving to the lower bin, and 11.0 \% moving to the higher bin. The sources that moved from the 0--1 bin to the higher 1--2 bin could be of special interest, because the nearest AGNs may be astrometrically perturbed in Gaia DR3. These, however, constitute only 3\% of the entire amount of misplaced sources. The entire process of proper motion field analysis was repeated for the redshift-classified samples, including transformation of the input astrometric data to Galactic coordinates, averaging of proper motions in trapezoidal cells and formal covariance computation, setting up the 3D condition matrices and 2D weights, computing the weighted normal equations, and least-squares solution for the 30 VSH coefficients. The results (for each $z$-bin) were compared to the main results in Table 1. The computed VSH coefficients are very close across the entire set of 30 harmonics. For example, for the most populous 1--2 bin, the statistically most significant terms \{mag,0,1,0\} and \{ele,1,2,2\} changed from 12.40 and 7.16 \uasyr\ to 12.66 and 7.38 \uasyr, respectively. The same level of consistency is found for the higher bins 2--3 and $>3$, as well as the overall sample.

\begin{deluxetable*}{cr|cc|cc|cc|cc} \label{qvsh.tab}
\tablehead{
\multicolumn{2}{c|}{} & \multicolumn{2}{c|}{$z>1$} &\multicolumn{2}{c|}{$1<z<2$} &
\multicolumn{2}{c|}{$2<z<3$} &\multicolumn{2}{c}{$z>3$} \\ \hline
\multicolumn{1}{c}{number} & \multicolumn{1}{c|}{VSH}  & \multicolumn{1}{c}{Value} & \multicolumn{1}{c|}{S/N}
& \multicolumn{1}{c}{Value} & \multicolumn{1}{c|}{S/N} & \multicolumn{1}{c}{Value} & \multicolumn{1}{c|}{S/N} & 
\colhead{Value} & \colhead{S/N}\\
\colhead{}   &  \multicolumn{1}{c|}{}   &   \colhead{$\mu$as yr$^{-1}$} & \multicolumn{1}{c|}{} & \colhead{$\mu$as yr$^{-1}$} & \multicolumn{1}{c|}{} & \colhead{$\mu$as yr$^{-1}$} &\multicolumn{1}{c|}{} & \colhead{$\mu$as yr$^{-1}$} & 
 \colhead{}  
 } 
\startdata
 1 & \{\text{mag},0,1,0\} & 13.00 & 12.79 & 13.16 & 11.06 & 13.09 & 6.37 & -26.30 & 1.84 \\
 2 & \{\text{mag},1,1,1\} & 0.99 & 1.02 & -2.37 & 2.05 & 9.22 & 4.86 & 21.38 & 2.40 \\
 3 & \{\text{mag},2,1,1\} & -5.49 & 4.19 & -10.41 & 6.73 & 8.99 & 3.47 & 1.16 & 0.08 \\
 4 & \{\text{ele},0,1\} & 0.41 & 0.42 & -1.99 & 1.72 & 8.26 & 4.13 & -30.20 & 2.25 \\
 5 & \{\text{ele},1,1,1\} & -15.99 & 13.75 & -15.11 & 11.01 & -18.98 & 8.25 & -22.75 & 1.74 \\
 6 & \{\text{ele},2,1,1\} & -2.07 & 2.09 & -2.91 & 2.47 & 0.35 & 0.18 & 1.15 & 0.15 \\
 7 & \{\text{mag},0,2,0\} & -0.94 & 2.29 & -1.02 & 2.09 & -1.07 & 1.35 & 4.06 & 1.22 \\
 8 & \{\text{mag},1,2,1\} & -3.48 & 5.85 & -3.41 & 4.82 & -3.98 & 3.45 & 4.53 & 1.01 \\
 9 & \{\text{mag},2,2,1\} & 1.26 & 1.96 & 1.28 & 1.68 & 1.31 & 1.03 & -3.09 & 0.44 \\
 10 & \{\text{ele},0,2\} & -1.09 & 2.85 & -1.20 & 2.63 & -0.79 & 1.07 & -2.42 & 0.79 \\
 11 & \{\text{ele},1,2,1\} & -1.61 & 2.24 & -2.05 & 2.40 & 0.12 & 0.08 & 2.93 & 0.38 \\
 12 & \{\text{ele},2,2,1\} & -1.19 & 1.99 & -0.85 & 1.19 & -2.43 & 2.13 & -3.00 & 0.66 \\
 13 & \{\text{mag},1,2,2\} & 1.63 & 2.61 & 1.30 & 1.76 & 2.26 & 1.85 & -6.15 & 1.03 \\
 14 & \{\text{mag},2,2,2\} & -0.30 & 0.38 & -1.69 & 1.81 & 3.02 & 1.90 & -4.06 & 0.44 \\
 15 & \{\text{ele},1,2,2\} & 7.75 & 10.42 & 7.56 & 8.64 & 8.24 & 5.55 & 11.81 & 1.26 \\
 16 & \{\text{ele},2,2,2\} & -1.29 & 1.98 & -0.70 & 0.91 & -4.51 & 3.56 & -0.47 & 0.08 \\
 17 & \{\text{mag},0,3,0\} & -0.20 & 0.57 & -0.03 & 0.06 & -1.09 & 1.59 & 13.16 & 3.79 \\
 18 & \{\text{mag},1,3,1\} & -4.94 & 12.21 & -4.99 & 10.40 & -4.65 & 5.88 & -10.66 & 3.03 \\
 19 & \{\text{mag},2,3,1\} & 0.95 & 2.37 & 0.59 & 1.24 & 1.99 & 2.53 & -3.03 & 0.84 \\
 20 & \{\text{ele},0,3\} & -0.97 & 3.06 & -0.59 & 1.56 & -2.01 & 3.19 & 3.07 & 0.97 \\
 21 & \{\text{ele},1,3,1\} & -0.11 & 0.28 & -0.17 & 0.35 & -0.25 & 0.32 & 7.19 & 1.97 \\
 22 & \{\text{ele},2,3,1\} & 0.21 & 0.55 & 0.31 & 0.69 & -0.22 & 0.29 & 1.59 & 0.52 \\
 23 & \{\text{mag},1,3,2\} & 2.32 & 5.58 & 2.16 & 4.37 & 2.54 & 3.14 & 3.22 & 1.00 \\
 24 & \{\text{mag},2,3,2\} & 1.44 & 2.98 & 1.27 & 2.22 & 2.22 & 2.32 & -1.19 & 0.23 \\
 25 & \{\text{ele},1,3,2\} & -1.12 & 2.24 & -1.64 & 2.77 & 0.03 & 0.03 & -4.87 & 0.94 \\
 26 & \{\text{ele},2,3,2\} & 0.95 & 2.23 & 1.31 & 2.57 & 0.08 & 0.09 & 2.57 & 0.76 \\
 27 & \{\text{mag},1,3,3\} & -0.93 & 1.80 & -1.46 & 2.41 & 0.36 & 0.36 & 4.35 & 0.85 \\
 28 & \{\text{mag},2,3,3\} & -0.94 & 1.79 & -0.82 & 1.31 & -1.07 & 1.02 & -10.37 & 1.94 \\
 29 & \{\text{ele},1,3,3\} & -1.75 & 3.42 & -1.84 & 3.05 & -1.38 & 1.36 & -7.44 & 1.39 \\
 30 & \{\text{ele},2,3,3\} & 1.22 & 2.38 & 0.67 & 1.11 & 2.35 & 2.35 & -3.98 & 0.86 \\
\enddata
\caption{VSH fits of Quaia/CRF proper motion fields.}
\end{deluxetable*}

\end{document}